\documentclass[a4paper]{aa} 

\usepackage{txfonts}
\usepackage{graphicx}
\usepackage{amssymb}
\usepackage{lscape}
\usepackage{natbib}
\usepackage{multirow}
\usepackage{supertabular}
\bibpunct{(}{)}{;}{a}{}{,} % to follow the A&A style

\def\feh {[Fe/H]}
\def\logg {\log g}

\def\ds {$\delta$ Sct}
\def\dss {$\delta$ Sct stars}

\def\corot {{CoRoT}}
\def\cesam {{\sc{cesam}}}
\def\graco {{\sc{graco}}}
\def\filou {{\sc{filou}}}

\def\period {{\sc Period04}}
\def\sigspec {{\sc SigSpec}}

\def\kepler {{\it{Kepler}}}
\def\stara {HD\,174936}
\def\starb {HD\,174966}
\def\strom {Str\"{o}mgren}
\def\logg {$\log g$}

\def\vsini {$\mathrm{v}\cdot\!\sin\!~i$}
\def\kms {$\mathrm{km}~\mathrm{s}^{-1}$}
\def\teff {$T_{\mathrm{eff}}$}
\def\msol {$\mathrm{M}_{\odot}$}
\def\rsol {$\mathrm{R}_{\odot}$}
\def\lsol {$\mathrm{L}_{\odot}$}
\def\muhz {$\mu\mbox{Hz}$}
\def\hc {$X_{\mathrm{c}}$}
\def\cd {$\mbox{d}^{-1}$}
\def\gcm {$\mathrm{g~cm^{-3}}$}
\def\Dnu {$\Delta\nu$}
\def\rhom {$\bar\rho$}
\def\alfa {$\alpha_{ML}$}
\def\ov {$\mathrm{d}_{ov}$}

\def\ele{$\ell$}

\begin{document}

  \title{An in-depth study of \starb\ with \corot\thanks{The CoRoT space mission was developed and is operated by the French space agency CNES, with participation of ESA's RSSD and Science Programmes, Austria, Belgium, Brazil, Germany, and Spain.} photometry and HARPS\thanks{This work is based on ground-based observations made with the ESO 3.6m-telescope at La Silla Observatory under the ESO Large Programme LP182.D-0356, and on observations collected at the Centro Astron\'{o}mico Hispano Alem\'{a}n (CAHA) at Calar Alto, operated jointly by the Max-Planck Institut f\"{u}r Astronomie and the Instituto de Astrof\'{\i}sica de Andaluc\'{\i}a (CSIC), and on observations made at Observatoire de Haute Provence (CNRS), France, and at Observatorio de Sierra Nevada (OSN), Spain, operated by the Instituto de Astrof\'{\i}sica de Andaluc\'{\i}a (CSIC). This research has made use of both the Simbad database, operated at CDS, Strasbourg, France, and the Astrophysics Data System, provided by NASA, USA.} spectroscopy. Large separation as a new observable for \dss.}
  \titlerunning{An in-depth study of \starb\ with \corot\ and HARPS}
  \authorrunning{Garc\'ia Hern\'andez et al.}

  \author{A. Garc\'ia Hern\'andez\inst{1,2}
  \and A. Moya\inst{3}
  \and E. Michel\inst{4}
  \and J. C. Su\'arez\inst{2}
  \and E. Poretti\inst{5}
  \and S. Mart\'in-Ru\'iz\inst{2}
  \and P. J. Amado\inst{2}
  \and R. Garrido\inst{2}
  \and E. Rodr\'iguez\inst{2}
  \and M. Rainer\inst{5}
  \and K. Uytterhoeven\inst{6,7}
  \and C. Rodrigo\inst{3,8}
  \and E. Solano\inst{3,8}
  \and J. R. Rod\'on\inst{2}
  \and P. Mathias\inst{9,10}
  \and A. Rolland\inst{2}
  \and M. Auvergne\inst{4}
  \and A. Baglin\inst{4}
  \and F. Baudin\inst{11}
  \and C. Catala\inst{4}
  \and R. Samadi\inst{4}
 }

  \offprints{A. Garc\'ia Hern\'andez, \email{agh@astro.up.pt}}

  \institute{Centro de Astrof\'{\i}sica, Universidade do Porto, Rua das Estrelas, 4150-762 Porto, Portugal \and Instituto de Astrof\'{\i}sica de Andaluc\'{\i}a (CSIC), CP3004, Granada, Spain \and Departamento de Astrof\'{\i}sica, Centro de Astrobiolog\'{\i}a (INTA-CSIC), PO BOX 78, 28691 Villanueva de la Ca\~nada, Madrid, Spain \and LESIA, Observatoire de Paris, CNRS UMR 8109, Universit\'{e} Pierre et Marie Curie, Universit\'{e} Denis Diderot, 5 place J. Janssen, 92195 Meudon, France \and INAF-Osservatorio Astronomico di Brera, Via E. Bianchi 46, I-23807 Merate (LC), Italy \and Instituto de Astrof\'{\i}sica de Canarias, 38200 La Laguna, Tenerife, Spain \and Departamento de Astrof\'{\i}sica, Universidad de La Laguna, 38205 La Laguna, Tenerife, Spain \and Spanish Virtual Observatory \and Universit\'e de Toulouse; UPS-OMP; IRAP; Tarbes, France \and CNRS; IRAP; 57, Avenue d'Azereix, F-65000 Tarbes, France \and Institut d'Astrophysique Spatiale, CNRS/Universit\'{e} Paris XI UMR 8617, F-091405 Orsay, France}
  \date{Received ... / Accepted ...}

  \abstract {}{The aim of this work was to use a multi-approach technique to derive the most accurate values possible of the physical parameters of the $\delta$ Sct star \starb, observed with the \corot\ satellite. In addition, we searched for a periodic pattern in the frequency spectra with the goal of using it to determine the mean density of the star.}
{First, we extracted the frequency content from the \corot\ light curve. Then, we derived the physical parameters of \starb\ and carried a mode identification out from the spectroscopic and photometric observations. We used this information to look for the models fulfilling all the conditions and discussed the inaccuracies of the method because of the rotation effects. In a final step, we searched for patterns in the frequency set using a Fourier transform, discussed its origin and studied the possibility of using the periodicity to obtain information about the physical parameters of the star.}{A total of 185 peaks were obtained from the Fourier analysis of the \corot\ light curve, being almost all reliable pulsating frequencies. From the spectroscopic observations, 18 oscillation modes were detected and identified, and the inclination angle ($62.5^{\circ}$$^{+7.5}_{-17.5}$) and the rotational velocity of the star (142~\kms) were estimated. From the multi-colour photometric observations, only 3 frequencies were detected, which correspond to the main ones in  the \corot\ light curve. We looked for periodicities within the 185 frequencies and found a quasiperiodic pattern \Dnu$\sim$64~\muhz. Using the inclination angle, the rotational velocity and an Echelle diagram (showing a double comb outside the asymptotic regime), we concluded that the periodicity corresponds to a large separation structure. The quasiperiodic pattern allowed us to discriminate models from a grid, finding that the value of the mean density is achieved with a 6\%\ uncertainty. So, the \Dnu\ pattern could be used as a new observable for A-F type stars.}{}


\maketitle

%++++++++++++++++++++++++++++++++++++++++++++++++++++++++++++++++++++++++++++++%
%                             Body of Paper                                    %
%++++++++++++++++++++++++++++++++++++++++++++++++++++++++++++++++++++++++++++++%

%--------------------------------------------------------------------------------------------------%
\section{Introduction\label{sec:intro}}
%--------------------------------------------------------------------------------------------------%

The asteroseismic interest of $\delta$ Scuti stars has progressively grown since it became evident that the detection of excited modes was limited by observational constraints, such as the duty cycle and the spectral window. \citet{garrido2004} showed how the number of detected frequencies increased with the refinement of the observational effort, i.e., a step-by-step process from single-site sporadic runs to multi-colour multisite campaigns. Since the multisite campaign on FG Vir revealed 75 frequencies \citep{breger2005}, it was predicted that hundreds of excited modes could be detected from the space monitoring of $\delta$ Sct stars.\par

CoRoT \citep[COnvection, ROtation and planetary Transits; ][]{baglin2006} and {\it Kepler} \citep{Borucki2010} space missions confirmed this prediction. In particular, several hundreds of frequencies were detected in the light curves of the CoRoT $\delta$ Sct stars HD~50844 \citep{poretti2009}, \stara\ \citep[][hereafter GH09]{garciahernandez2009} and HD~50870 \citep{mantegazza2012}. The debate on their pulsational nature is still open \citep{mantegazza2012} since other atmospheric effects such as the granulation \citep{kallinger2010} have been invoked to explain the very rich amplitude spectrum. Nonetheless, pulsating stars seem to have enough energy to excite such a high number of modes \citep{moyacris2010}. Results on {\it Kepler} \dss\ are described by \citet{grigahcene2010}, \citet{katrien2011} and \citet{balonadziembowski2011}.\par

The usual approach to studying \dss\ is through frequency analysis and mode identification. Other methods have been explored to extract more information from the observations. The search for regularities in the Fourier spectrum is one of the methods typically used to study the frequency spectra of stars whose modes are in the asymptotic regime, as in the case for the Sun and solar-like pulsators. \ds\ modes are generally located near the fundamental radial mode and outside the asymptotic regime, so regularities in their frequency sets were not expected. However, regular spacing have been claimed for some \dss\ \citep{handler1997, Breger99}. These works employed two different methods to look for regularities: through the calculation of the Fourier transform of the frequency set \citep{handler1997}, and through a histogram of the frequency differences \citep{Breger99}. For a few stars only, they made marginal detections of periodic structures. Examples of \dss\ with a high number of modes are needed to confirm that such spacings are usually present in this type of pulsators.\par

GH09 used the high number of frequencies obtained from \corot\ data to find periodicities in the frequency spectrum of \stara. They detected a periodic pattern and pointed out that this periodicity seems to be caused by a large separation structure, following its definition as $\Delta\nu_{\ell}=\nu_{n+1,\ell}-\nu_{n,\ell}$, being $n$ the radial order of the mode and \ele, its spherical degree. This method was successfully used to find patterns in other \dss\ observed from space \citep{mantegazza2012,garciahernandez2013}. Nonetheless, in absence of information about the inclination angle of the star, we could not rule out rotational splitting as the origin behind the pattern.\par
 
In the work we present here, we carried out an asteroseismic analysis of another \ds\ star observed by \corot, \starb\ (Sec.~\ref{sec:corot}), confirming that a pattern structure in the detected frequencies is present. In addition, we performed new spectroscopic (Sec.~\ref{sec:spectroscopy}) and \strom\ photometric (Sec.~\ref{sec:stromgren}) observations. We derived the star's physical parameters, including the inclination angle and the rotational velocity, and we performed a mode identification. The identification allowed us to discriminate between models representative of the star (Sec.~\ref{sec:uncertainty}) and determine other physical quantities, such as the mass and the age. The analysis became a test for the theories of stellar interiors and oscillations.\par

Using the information on the inclination angle, the rotational velocity of \starb\ and from an Echelle diagram, we conclude that the most probable origin of the regular pattern presented in the frequency spectrum (Sec.~\ref{sec:patterns}) is a large separation structure. Finally, we analysed the possibility that the periodicity becomes a new asteroseismic observable for \dss\ (Sec.~\ref{sec:constrain}).\par

%--------------------------------------------------------------------------------------------------%
\section{Analysis of the light curve observed by \corot\label{sec:corot}}
%--------------------------------------------------------------------------------------------------%
The \ds\ variability of \starb\ was discovered in the preparatory work of the CoRoT mission \citep{Poretti_corot03}. \starb\ was observed in the same CoRoT frame as the other \ds-type star HD~174936 (GH09), during the first short run SRc01, between April 2007 and May 2007. The time span of the collected dataset was $\Delta$T = 27.2 days, with a sampling of one point every 32 sec. The final dataset consisted of 66481 data points after removing those points considered unreliable\footnote{Essentially points flagged by the reduction pipeline: hot pixels detected, the passing through the South Atlantic Anomaly (SAA), interpolated points, etc.}. The corresponding Rayleigh frequency resolution is (1/$\Delta$T) = 0.037\,\cd, and an oversampling of 20 corresponds to a frequency spacing of 0.0018 \cd. This is equivalent to the 
frequency spacing in mode ``High'' in the program package \period\ \citep{lenz2005}.\par

The data were  corrected for instrumental drift \citep{auvergne2009} by performing a linear fit to the light curves, and analysed with \period\ \citep{lenz2005} and \sigspec\ \citep{reegen2007}. The agreement between the two methods was excellent. This test was also successfully carried out for some other CoRoT targets such as HD~174936 (GH09) with similar time series, HD~50844 \citep{poretti2009} with $\Delta$T = 56.7 days and 140016 data points, HD~49434 \citep{chapellier2011} with $\Delta$T = 136.9 d and 331291 data points, and HD~50870 \citep{mantegazza2012} with $\Delta$T = 114.41 d and 307570 data points. In all the cases, \period\ was used to investigate the first 20, 200, and 500 peaks, respectively. \par

\begin{figure}
\begin{center}
 \includegraphics[angle=-90,width=0.5\textwidth]{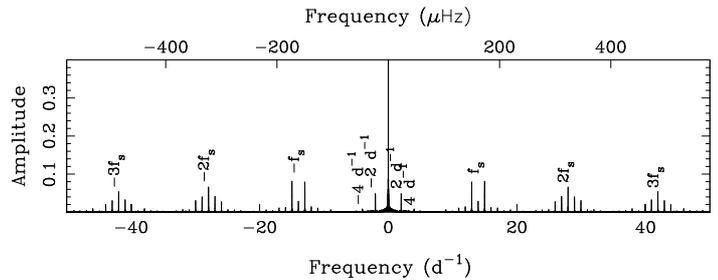}
 \caption{Spectral window of the short-run CoRoT dataset of \starb. $\mathrm{f}_\mathrm{s}$ is the orbital frequency of the satellite. Amplitude in the ordinate axis is normalized to the highest peak.}
 \label{fig:window}
\end{center}
\end{figure}

The spectral window associated to \starb\ (see Fig. \ref{fig:window}) is typical for all the targets observed by the CoRoT satellite in the same run. That is, the periodograms show neither the typical aliases at 1~\cd, nor the power levels that are common for ground-based data. On the contrary, all the aliases are related to effects produced by the satellite and its orbital frequency ($\mathrm{f}_\mathrm{s}$~=~13.972~\cd), and their power levels are much lower than those usual for ground-based data. In addition, high power peaks are also produced at 2.005 and 4.011~\cd, which come from the twice-daily SAA crossing.\par

%-fig-%
%-fig-%

To avoid problems with power close to the zero frequency, the analysis with \sigspec\ was carried out in the range 0.05-100~\cd. In the case of \period, the amplitude signal-to-noise ratio S/N~=~4.0 is the limit commonly used to consider a frequency as significant. In the case of \sigspec, the parameter used for significance is ``sig'' (= spectral significance), and the default limit is sig = 5.0. This is equivalent to about S/N = 3.8 \citep[and sig = 5.46 is approximately equivalent to S/N = 4.0;][]{reegen2007, kallinger2008}. However, we used a much more conservative limit, namely sig = 10.0, because the corresponding S/N values, determined using \period\ on the residuals, were much lower than expected\footnote{The S/N obtained on the residuals for a sig = 5.0 was lower than the expected value S/N = 3.8. We used the limit sig = 10.0 to assure that we are above the usual limit of S/N = 4.0.}. This is probably caused by a high number of peaks still remaining among the residuals in the region of interest. This was explained in much more detail in similar recent works for other CoRoT targets: HD\,50844 \citep{poretti2009}, HD\,49434 \citep{chapellier2011} and HD\,50870 \citep{mantegazza2012}.\par

The limit of sig = 10 was achieved after removing 185 peaks. This corresponded to a level of about 8~ppm for the smallest amplitudes. Table~\ref{tab:frequencies} (in the appendix) lists the frequencies obtained along with the most relevant parameters. The S/N values were calculated using \period\ on the residual file provided by the \sigspec\ package. Each S/N value was calculated within a box of width 5~\cd\ centred on the corresponding peak, as is usual for this parameter \citep{breger1993,rodriguez2006}. In addition, columns nine to eleven list the {\emph{formal}} error bars for frequencies, amplitudes and phases determined using the formulae by \citet{montgomery1999}.\par

The last column of the table lists possible identifications of harmonics or combinations as well as some controversial peaks, which appeared closer, within the frequency resolution, to another peak with higher amplitude. The origin of these controversial peaks could be related to a non-precise pre-whitening of a frequency from the light curve, generating spurious detections in the following analysis of the residuals.\par

We investigated harmonics and combinations up to the third order between the main peaks within a range of $\pm\ $0.010~\cd. This study has been made taking into account only the first 12 main peaks of the list (F1 to F12), harmonics until the fifth for F1 to F4 and combinations considering the first, second and third orders (i.e., $AF_{a}+BF_{b}$ being $A$, $B = [1,3]$). The possibility of interactions with the satellite orbital frequency (assumed as f$_s=13.972$~\cd, FS in the table) has also been studied for the first 4 main peaks (F1 to F4) and four first harmonics of FS (FS to 4FS). We have also studied the possibility of remaining peaks (residuals) corresponding to the sidelobes of 1~\cd\ alias around FS and its harmonics.\par

A total of 37 possible combinations and harmonics were detected. However, most of them seemed to be coincidences. We investigated how good a frequency matched its exact {\emph{theoretical}} combination. If a frequency, let's say $F_{ab}$, is the result of a non-linear interaction between two (or more) frequencies, $F_{a}$ and $F_{b}$ for example, then its value should be exactly (theoretically speaking) the linear combination of the parent frequencies: $F_{ab} = F_{a+b} \equiv AF_{a} + BF_{b}$ \citep{garrido1996}. We used the uncertainties in the peak and in the parent frequencies to check whether a frequency could be a combination. If the difference between the frequency and its mathematical combination is greater than the sum of all errors\footnote{To derive uncertainties in the determination of mathematical combinations, we follow the standard theory on error propagation: $|\sigma F_{a+b}| = |A\cdot\sigma F_{a}| + |B\cdot\sigma F_{b}|$. Because no information about the uncertainty in the satellite orbital frequency is available, FS was taken as a constant.} within 1$\sigma$, i. e. $|F_{ab} - F_{a+b}| > \sigma F_{ab} + \sigma F_{a+b}$, then the probability of such frequency to be a combination is quite low. Finally, 25 frequencies fulfilled this condition and 19 of them still could be discarded as combinations when a 2$\sigma$ error instead was used in the operation. Therefore, only 12 peaks of the original 37 remained as possible combinations.\par

This result is quite different from the one obtained for the \corot\ target \stara\ (GH09), where no combination frequencies were reliably identified among the main frequency peaks. In fact, when comparing the periodograms of \starb\ and \stara\ we see two main differences: \starb\ shows only few high-amplitude peaks, with amplitudes higher than detected in \stara, and overall a smaller number of frequencies is detected (185 significant peaks versus 422 peaks for \stara).\par

\begin{figure}
\begin{center}
 \includegraphics[width=0.45\textwidth]{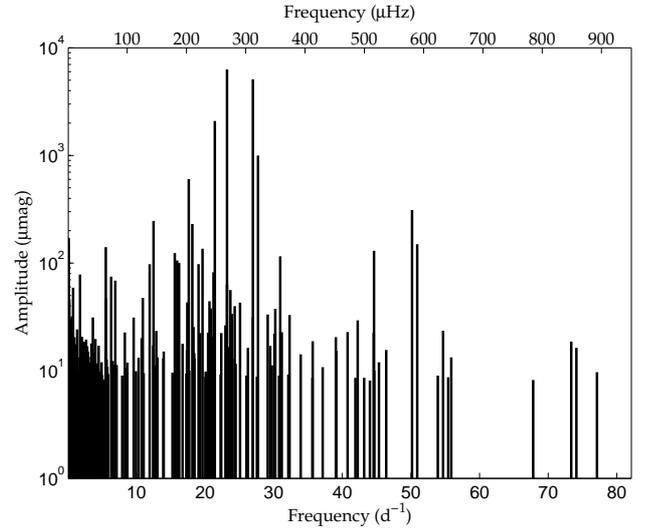}
 \caption{Bar plot of the 185 frequencies extracted from the \corot\ light curve of \starb. Note that the {\it{amplitude}} axis is in logarithmic scale and $\mu$mag units, and that the {\it{frequency}} is shown in \muhz\ and \cd\ units.}
 \label{fig:frequencies}
\end{center}
\end{figure}

The range of statistically significant detected frequencies for \starb\ goes from some value close to zero up to about 77~\cd, i.e. 900~$\mu$Hz (1~\cd~=~11.57~$\mu$Hz). But the highest amplitude peaks are grouped around 25~\cd\ (300~$\mu$HZ, see Fig. \ref{fig:frequencies}).\par

%-fig-%
%-fig-%

%--------------------------------------------------------------------------------------------------%
\section{Spectroscopic analysis\label{sec:spectroscopy}}
%--------------------------------------------------------------------------------------------------%

%--------------------------------------------------------------------------------------------------%
\subsection{Observations\label{subsec:spectroscopic observations}}
%--------------------------------------------------------------------------------------------------%

The spectroscopic observations were performed between June 12, 2009 and August 5, 2009 (not contemporary with the \corot\ data), covering a baseline of 53 days with 341 spectrograms, and were performed with three different high-resolution spectrographs, as reported in Table~\ref{tab:espectroscopy}.\par

%--table--
\begin{table}
 \caption{Specifications of the spectroscopic observation. The columns list the spectrograph used for each observation, the number of spectrograms (\#), the resolving power $R$, the number of nights, the observatory, and the exposure time, respectively.}
 \begin{center}
 \begin{tabular}{cccccc}
  \hline
  \hline
  Spect. & \# & $R$ & N$^{\circ}$ nights & Observatory & Exp. time (s) \\
  \hline
  HARPS & 104 & 80000 & 12 & ESO-La Silla & 1200 \\
  FOCES & 155 & 65000 & 21 & Calar Alto & 900 \\
  SOPHIE & 81 & 75000 & 9 & Haute Provence & 700 \\
  \hline
  \hline
 \end{tabular}
 \label{tab:espectroscopy}
 \end{center}
\end{table}
%--table--

The reduced spectrograms were brought to the same resolution of 65000 and then for each of them a mean profile was obtained by means of the Least-Squares Deconvolution (LSD) technique \citep{donati1997} with values between $-$300 and +300~\kms\ with a 2~\kms\ step. The spectral region between 4140 and 5670~$\AA$ was used when calculating the LSD profiles, taking care to omit the regions containing hydrogen lines. The median S/N ratios of these profiles, as computed from the dispersion of their adjacent continua, are 5640, 1004, and 2945 for the HARPS, FOCES and SOPHIE spectrograms, respectively.\par

\begin{figure}
\begin{center}
 \includegraphics[width=0.45\textwidth]{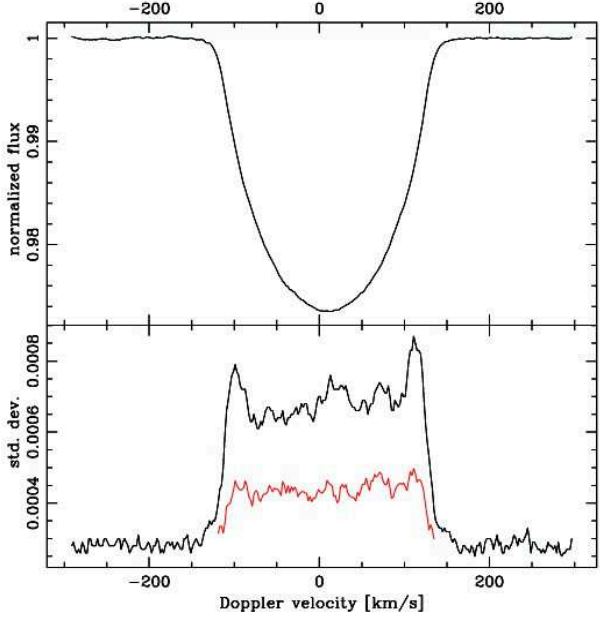}
 \caption{Average of the LSD profiles (upper panel). Lower panel: r.m.s. standard deviation of the individual LSD profile with respect to the average one (black); residual standard deviation after the fit of the detected terms (red).}
 \label{fig:profiles}
\end{center}
\end{figure}

The weighted average of the LSD profiles and its standard deviation are shown in the top and bottom panels, respectively, of Fig.~\ref{fig:profiles}. From the first three zeros of the Fourier transform of this average profile it has been possible to determine \vsini~$=126.1\pm1.2$ \kms.\par

%-fig-%
%-fig-%

%--------------------------------------------------------------------------------------------------%
\subsection{Physical parameters\label{subsec:spec_parameters}}
%--------------------------------------------------------------------------------------------------%

%-table-%
\begin{table}
\begin{center}
\caption{Physical parameters of \starb.}
\begin{tabular}{ccc}
\hline
\hline
Parameter & Value  & Reference \\
\hline
$<V>$    & 7.698~$\pm$~0.002 & 1 \\
$(b-y)_0$& 0.142~$\pm$~0.001 & 1 \\
$E_{b-y}$& 0.006~$\pm$~0.003 & 1 \\
$m_0$    & 0.187~$\pm$~0.002 & 1 \\
$c_0$    & 0.848~$\pm$~0.007 & 1 \\
$\beta$  & 2.796~$\pm$~0.001 & 1 \\ 
\vsini\  (\kms) & 126.1~$\pm$~1.2   & 2 \\
\teff\ (K) & 7555~$\pm$~50 & 2 \\
\logg    & 4.21~$\pm$~0.05 & 2 \\
\feh     &$-0.08~\pm$~0.10 & 2 \\
M  (\msol)   & 1.70~$\pm$~0.20 & 2 \\
R (\rsol)   & 1.70~$\pm$~0.20 & 2 \\
L(R,T) (\lsol)    &  8.5~$\pm$~2.0 & 2 \\
\hline
   \end{tabular}
   \tablebib{(1) GAUDI archive \citep{solano2005}; (2) this work.}
   \label{tab:parfis}
 \end{center}
\end{table}

The Ground-Based Seismology Working Group \citep{catala2006} prepared the GAUDI archive\footnote{http://sdc.cab.inta-csic.es/gaudi/} \citep{solano2005} for the CoRoT mission, collecting high-resolution spectroscopy and multicolour Str\"omgren photometry. The analysis of the $uvby\beta$ photometry supplied the (preliminary) parameters: \teff~=~$7637\pm200$~K, \logg~=~$4.03\pm0.20$~dex and \feh~=~$-0.11\pm0.20$~dex \citep{MoonDwo85}; very useful in the exploratory work for target selection, but the subsequent HARPS monitoring allowed us to perform a more accurate analysis.\par

The spectroscopic analysis was performed by using the set of HARPS spectra rather than the single spectrum taken with ELODIE at Observatoire Haute Provence available in the GAUDI archive. To this end, non-linear least-squares fits were performed on some regions of the very high S/N mean stellar spectrum by means of the SME code \citep[][]{valenti1996}. In particular, the regions around H$\alpha$, H$\beta$, H$\gamma$, MgI 5180 triplet, 5300-5340 $\AA$, and 4560-4795 $\AA$ were separately fitted. {{The \teff, \logg, and \feh\, values thus obtained are listed in Table~\ref{tab:parfis} with their internal errors}}. These HARPS values agree with those supplied by Str\"omgren photometry and with those from the ELODIE spectrum (both from the GAUDI archive) within the respective error bars.\par

\begin{figure}
\begin{center}
 \includegraphics[width=0.50\textwidth]{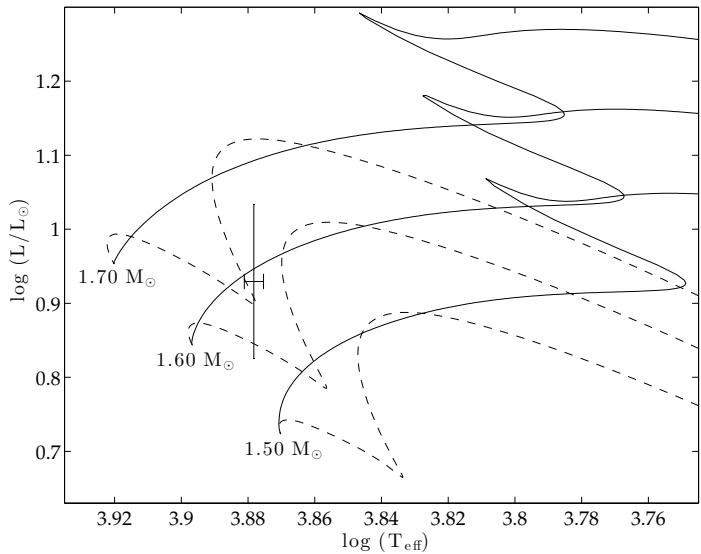}
 \caption{HR diagram showing evolutionary tracks for 1.5, 1.6 and 1.7~\msol. Models were computed using \feh$=-0.08$~dex, \alfa$=0.5$ and \ov$=0.2$ from the pre-main sequence phase (dashed line) to the end of the main sequence (solid line). They included OPAL opacity tables and Eddington atmospheres. The position of \starb\ and the spectroscopic $1\sigma$ uncertainty box derived in this work are also shown (see Table~\ref{tab:parfis}).}
 \label{fig:hr}
\end{center}
\end{figure}

The \teff~=~7555~K value was used to compute the bolometric correction BC~=~0.033~mag by means of the formulae given by \citet{Torres2010}. Str\"omgren photometry from GAUDI archive allowed us to determine the colour excess $E_{b-y}$ and in turn the interstellar absorption $A_V$~=~0.025 mag. These corrections are very small with respect to the uncertainty on the HIPPARCOS parallax, {{7.81~$\pm$~0.66~mas \citep{vanleeuwen2007}; from which $M_{\rm bol} = 2.16\pm0.18$ and L~=~10.7~$\pm$~1.7~\lsol\ could be determined.}} From M$_{bol}$ and \teff, a value of R~=~$1.90\pm0.19$~\rsol\ can be obtained and, using the \logg\, value, M~=~$2.16\pm0.50$~\msol. This mass is higher than that calculated using the evolutionary tracks \citep[M~=~$1.68\pm0.1$~\msol; ][]{Scha92} and the  $uvby-$H$_{\beta}$ calibration \cite[M~=~$1.82\pm0.09$~\msol; ][]{ribas1997}, but within the error bars derived from the calibration mean errors. We finally adopted the self-consistent values listed in Table~\ref{tab:parfis} for the subsequent analysis of the line profile variations. {{An HR diagram showing a model evolutionary track with those parameters adopted is plotted in Fig.~\ref{fig:hr}.}} We also estimated that the fundamental radial mode frequency is $17.3 \pm 2.5$~\cd\ \citep[Eq.~6 in ][]{Breger00} and we noted that this value is quite close to F5~=~17.62~\cd\ in the \corot\ frequency list (Table~\ref{tab:frequencies}).\par

%-fig-%
%-fig-%

Before proceeding to the mode identification we derived the average line parameters by fitting the mean profile with the FAMIAS software \citep{zima2008}. We found a barycentric velocity of $9.1\pm0.5$~\kms, an intrinsic line width of $8.1\pm1$~\kms\, and \vsini~=~$126.2\pm1.2$~\kms. The latter value is  in perfect agreement with that obtained from the Fourier transform. Assuming R~=~1.8~\rsol, this value corresponds to a break-up rotational velocity of 424~\kms,  which supplied the important constraint $i>17.5^{\circ}$.\par

%--------------------------------------------------------------------------------------------------%
\subsection{Line profile variations (LPV)\label{subsec:lpv}}
%--------------------------------------------------------------------------------------------------%

We analysed the line-profile variations of the time series consisting of 341 LSD profiles using the pixel-by-pixel Least-Squares approach described by \cite{mantegazza2000} and by the pixel-by-pixel Fourier technique \citep{zima2006}. In the analysis we considered the part of the profile between $-$119 and +135 \kms.\par

Among the detected terms there were three low-frequency ones (0.039, 0.253, and 0.472~\cd) and three high-frequency ones (46.96, 58.73, and 59.81~\cd). All these terms do not have photometric counterparts in the \corot\ lightcurve. Moreover, the behaviour of amplitude and phase curves across the line profile of the low-frequency terms are not easily interpretable in terms of pulsation modes, and we suspect that they were introduced in the time series by the merging of data from three different instruments. The three high-frequency peaks have the appearance of low-degree retrograde modes. Taking into account  their non-photometric detection, we suspect that they are aliases above the pseudo-Nyquist frequency of undetected lower frequency modes (in the range 20-40~\cd). Since we cannot be sure of their physical significance, we did not perform any mode identification for these six terms.\par

%--table--
\begin{table*}
 \caption{Table with the results of the mode identification from spectroscopic and \strom\ observations, and the values of $n$ and \ele\ obtained in the model fitting (last two columns). The spectroscopic amplitudes are in continuum units as defined by \citet{zima2006}. Columns 8 and 9 show the corresponding \corot\ photometric frequencies. The spectroscopic frequencies 27.709 and 27.720~\cd\ are uncertain.}
 \begin{center}
 \begin{tabular}{cccccccr@{ = }lccc}
  \hline
  \hline
    \multicolumn{5}{c}{Spectroscopy} & \multicolumn{2}{c}{\strom\ photometry} & \multicolumn{3}{c}{\corot\ photometry} & \multicolumn{2}{c}{Model fit}\\
    Freq. (\cd\ --- \muhz) & Ampl. & S/N & \ele\ ($\pm1$) & $m (\pm1)$  & Freq. (\cd) & \ele\ & \multicolumn{2}{c}{Freq. (\cd)} & Ampl. (mmag) & \ele & $n$ \\
   \hline
    6.801 --- 78.72 & 0.033 & 6.5 & 15 & -15 & --- & --- & \multicolumn{2}{c}{---} & --- & --- & --- \\
    18.131 --- 209.85 & 0.023 & 3.8 & 3 & 1 & --- & --- & F8 & 18.135 & 0.231 & 3 & -2 \\
    21.427 --- 248.00 & 0.031 & 4.6 & 1 & 1 & 21.421 & 2, 3, 1\&3 & F3 & 21.420 & 2.093 & --- & --- \\
    23.152 --- 267.96 & 0.044 & 8.2 & 1 & -1 & --- & --- & \multicolumn{2}{c}{---} & --- & 2 & 0 \\
    23.192 --- 268.43 & 0.056 & 13.0 & 1 & -1 & 23.195 & 0, 1 & F1 & 23.195 & 6.290 & 0 & 2 \\
    24.122 --- 279.19 & 0.028 & 5.3 & 1 & -1 & --- & --- & \multicolumn{2}{c}{---} & --- & 1 & 1 \\
    25.023 --- 289.62 & 0.035 & 7.0 & 7 & 5 & --- & --- & \multicolumn{2}{c}{---} & --- & --- & --- \\
    26.025 --- 301.22 & 0.042 & 11.0 & 8 & 8 & --- & --- & {{F171}} & {26.026} & {0.0090} & --- & --- \\
    26.955 --- 311.98 & 0.041 & 7.9 & 2 & 1 & 26.956 & 0, 1, 2 & F2 & 26.958 & 5.103 & 2 & 1 \\
    26.982 --- 312.29 & 0.033 & 7.7 & 7 & 3 & --- & --- & {F46} & {26.962} & {0.0312} & --- & --- \\
    27.535 --- 318.69 & 0.028 & 6.4 & 8 & 4 & --- & --- & {F172} & {27.543} & {0.0088} & --- & --- \\
    27.709 --- 320.71 & 0.042 & 10.7 & 3 & 3 & --- & --- & F4 & 27.715 & 0.997 & --- & --- \\
    27.720 --- 320.83 & 0.048 & 11.3 & 2 & 2 & --- & --- & F4 & 27.715 & 0.997 & --- & --- \\
    30.093 --- 348.30 & 0.027 & 4.9 & 2 & 2 & --- & --- & {F67} & {30.101} & {0.0221} & --- & --- \\
    30.954 --- 358.26 & 0.028 & 5.6 & 7 & 5 & --- & --- & F17 & 30.950 & 0.116 & --- & --- \\
    31.171 --- 360.78 & 0.044 & 8.9 & 10 & 6 & --- & --- & F61 & 31.178 & 0.023 & --- & --- \\
    32.026 --- 370.67 & 0.032 & 7.7 & 10 & 5 & --- & --- & \multicolumn{2}{c}{---} & --- & --- & --- \\
    32.561 --- 376.86 & 0.041 & 7.9 & 11 & 9 & --- & --- & \multicolumn{2}{c}{---} & --- & --- & --- \\
\noalign{\smallskip}
  \hline
 \end{tabular}
 \label{tab:lpv}
 \end{center}
\end{table*}
%--table--

The mode identification has been performed by means of the FAMIAS code \citep{zima2008}. Table~\ref{tab:lpv} lists the 18 detected modes in order of increasing frequency. The second column gives their amplitudes, in continuum units as defined in the FAMIAS software. These amplitudes are a measurement of the contribution of each term to the whole line profile variability. We report the S/N ratio, always as defined in FAMIAS, in the third column, and the best fitting \ele\ and $m$ values in the fourth and fifth columns (negative $m$ values correspond to retrograde modes).\par

Some of the spectroscopically-detected modes have a photometric counterpart. In this case photometric frequencies and amplitudes are reported in the sixth and seventh columns. In particular, the frequency analysis of the Line Profile Variations showed a distorted peak at the fourth photometric term (27.715~\cd). It is probably an unresolved bunch of peaks. We could tentatively disentangle it as a doublet composed of two modes at 27.720 and 27.709~\cd, but being at the limit of the spectroscopic resolution their values remain uncertain.\par

\begin{figure}
\begin{center}
 \includegraphics[width=0.45\textwidth]{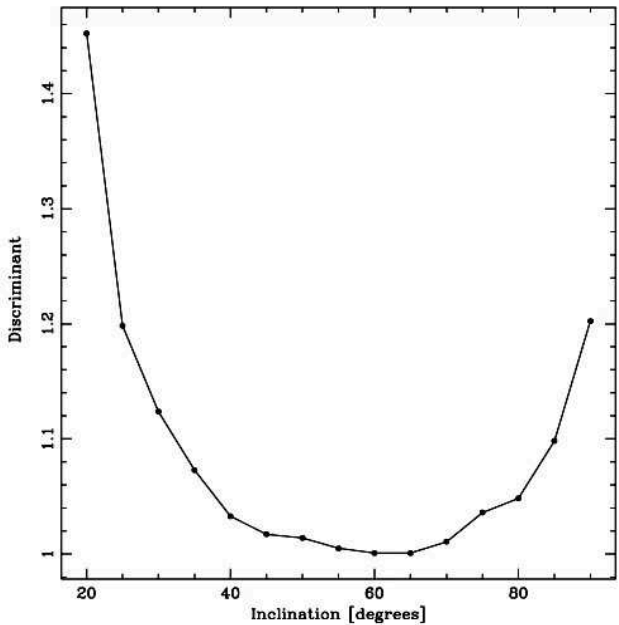}
 \caption{Discriminant supplied by the simultaneous fit of all the identified modes versus the rotational axis inclination.}
 \label{fig:inclination}
\end{center}
\end{figure}

For each mode in Table~\ref{tab:lpv}, a best fit was performed on its amplitude and phase behaviours across the line profile, keeping as free parameters: the velocity, the spherical degree, the azimuthal order, the amplitude, the phase, and the inclination. Then all the modes were fitted together leaving for each of them the velocity amplitude as a free parameter and changing each time the inclination with a step of 5$^{\circ}$ from 20$^{\circ}$ to 90$^{\circ}$. In such a way we could obtain a discriminant value for each inclination angle, $i$, by computing the differences between the observed and computed line-profile variations, as  described in \citet{Mantegazza2000_conf}. {{The case of \starb\, is less favourable than others (e.g., \citealt{mantegazza2000}). Indeed, the behaviour of the discriminant shows a broad minimum centered at 62.5$^{\circ}$, with very similar values in the interval from 45$^{\circ}$ to 70$^{\circ}$ (Fig.~\ref{fig:inclination}). For such an  interval the stellar equatorial rotational velocities are within a well-constrained range, i.e., 135--178~\kms. The corresponding rotational frequencies are in the 1.48--1.96~\cd\ (17--23~\muhz) range.\par

To proceed in our analysis we assumed $i = 62.5^{\circ}$, $v$~=~142~\kms\, and a rotational period of 0.64~d (rotational frequency 1.56~\cd~=~18.07~\muhz) and we checked again the \ele, $m$ values of each mode.}} Only slight changes were found for some of them, all within the uncertainties of the \ele, $m$ determinations which are at least of the order of the unity (\ele\ values are generally more reliable than $m$ ones, especially for high-m order modes). The uncertainties on $\ell, m$ values apply in particular to the close values of frequencies 23.152, 23.192, and 24.122 d$^{-1}$, which are identified by FAMIAS with the same $\ell, m$ couple. This could also be due to the limited frequency resolution of the spectroscopic time series, which prevents us to obtain clear amplitudes and phases behaviours across the line profile in case of close frequencies.\par

The case of the 6.8~\cd\ component is interesting from a methodological point of view. Its phase curve across the line profile is typical for a high-degree prograde mode. However, this is very unlikely as 6.8~\cd\ is much lower than the calculated fundamental radial frequency ($17.3\pm2.5$~\cd, see Sect.~\ref{subsec:spec_parameters}). The frequency of retrograde modes is lower in the observer's frame since they are travelling against the rotation. If the mode has a very-high $m$ order, the resulting frequency in the observer's frame is negative and we observe its mirrored peak in the positive plane. This could be the case for this component when considering $m\sim-15$ and the rotational frequency calculated above.\par

%-fig-%
%-fig-%

%--------------------------------------------------------------------------------------------------%
\section{Mode identification from \strom\ photometry\label{sec:stromgren}}
%--------------------------------------------------------------------------------------------------%

We obtained new multicolour photometric observations for \starb\ to perform a mode identification. The observations were made using the 0.9-m telescope at Observatorio de Sierra Nevada (OSN), Spain, by means of a \strom\ six-channel simultaneous photometer \citep{Nielsen83}. A total of 953 points were collected in the four $uvby$ bands simultaneously, during 29 nights in 2007, between May 15 and August 23, and 7 nights in 2008, between July 1 and 11. To do the differential photometry, HD\,173369 was selected as comparison star and HD\,181414 as check star. Both stars are close to our object and, to date, they do not present any sign of variability in our ground-based observations.\par

Frequency analyses were carried out using the program \period\ \citep{lenz2005}. The values of the frequencies found in the light curves of the four bands were the same. The signal-to-noise ratio reached in the $v$ and $b$ bands were higher than in the $u$ and $y$ bands. The $u$ band is the noisiest \citep{Martin00}. Table~\ref{tab:strom} shows the results of the amplitudes and phases for each frequency in the four bands. The values of the frequencies correspond to the highest amplitude frequencies detected in the \corot\ dataset.\par

%--table--
\begin{table*}
   \caption{Frequencies obtained from \strom\ photometry. The frequencies are consistent in all the bands.}
  \begin{center}
  \vspace{1em}
   \renewcommand{\arraystretch}{1.2}
   \begin{tabular}[ht!]{c|cccr@{=}@{$\pm$}}
   \hline
   \hline
   Band & Frequency (\cd) & Amplitude (mmag) & Phase (rad)  \\
     \hline
     \multirow{3}{*}{$u$} & $\mathrm{F}_{1}$ = 23.19481 $\pm$ 0.00004 & 8.2 $\pm$ 0.6 & 5.17 $\pm$ 0.08 \\
   				& $\mathrm{F}_{2}$ = 26.95847 $\pm$ 0.00004 & 6.5 $\pm$ 0.6 & 0.5 $\pm$ 0.1 \\
   				& $\mathrm{F}_{3}$ = 21.4209 $\pm$ 0.0001 & 2.7 $\pm$ 0.6 & 1.7 $\pm$ 0.2 \\
     \hline
     \multirow{3}{*}{$v$} & $\mathrm{F}_{1}$ = 23.19481 $\pm$ 0.00004 & 8.7 $\pm$ 0.3 & 4.96 $\pm$ 0.03 \\
   				& $\mathrm{F}_{2}$ = 26.95847 $\pm$ 0.00004 & 7.4 $\pm$ 0.3 & 0.24 $\pm$ 0.04 \\
   				& $\mathrm{F}_{3}$ = 21.4209 $\pm$ 0.0001 & 3.0 $\pm$ 0.3 & 1.1 $\pm$ 0.1 \\
     \hline
     \multirow{3}{*}{$b$} & $\mathrm{F}_{1}$ = 23.19481 $\pm$ 0.00004 & 7.8 $\pm$ 0.3 & 5.00 $\pm$ 0.04 \\
   				& $\mathrm{F}_{2}$ = 26.95847 $\pm$ 0.00004 & 6.8 $\pm$ 0.3 & 0.26 $\pm$ 0.04 \\
   				& $\mathrm{F}_{3}$ = 21.4209 $\pm$ 0.0001 & 2.4 $\pm$ 0.3 & 1.3 $\pm$ 0.1 \\
     \hline
     \multirow{3}{*}{$y$} & $\mathrm{F}_{1}$ = 23.19481 $\pm$ 0.00006 & 6.8 $\pm$ 0.4 & 5.02 $\pm$ 0.06 \\
   				& $\mathrm{F}_{2}$ = 26.95847 $\pm$ 0.00008 & 5.0 $\pm$ 0.4 & 0.29 $\pm$ 0.08 \\
   				& $\mathrm{F}_{3}$ = 21.4209 $\pm$ 0.0002 & 2.1 $\pm$ 0.4 & 1.5 $\pm$ 0.2 \\
     \hline
     \hline
   \end{tabular}
 \end{center}
 \label{tab:strom}
\end{table*}
%--table--

We used the amplitude and phase information of each frequency to identify the associated mode. The phase-shifts versus amplitude-ratios diagrams \citep{Garrido90} allow us to determine the spherical degree, \ele. To that end, we computed non-rotating equilibrium and non-adiabatic oscillation models. We used the evolutionary code {\cesam} \citep{Morel97,morel2008} and the pulsation code {\graco} \citep{Moya04,graco08} to calculate the equilibrium models and the non-adiabatic frequencies with \ele\ = [0, 3], respectively. Computations included realistic atmosphere models \citep{kurucz1993} to determine the non-adiabatic {{observables: $\phi^{T}$, ${\delta T_{eff}}/{T_{eff}}$ and ${\delta g_{e}}/{g_{e}}$.}} \par

Non-adiabatic processes are sensitive to convection \citep{dupret2003,pagoda03,Moya04}. The larger the path travelled by a convective cell, i. e., the mixing length distance, the more heat exchanged with its surroundings. We could then obtain information about the efficiency of the convection in \starb\ through the free parameters \alfa, of the mixing length theory, and \ov, where overshooting was used in every convection zone that appeared in the modelling, but not in the atmosphere. We used three values of each one in our computations: \alfa~=~0.5, 1.0 and 1.5, and \ov~=~0.1, 0.2 and 0.3, and compared the theoretical non-adiabatic observables of the models with the phase shifts and amplitude ratios observed. A main result is that models with \alfa~=~1.5 did not produce any phase shift vs. amplitude ratio diagrams compatible with the observations. This inefficiency in the convection for \dss\ was previously pointed out by \citet{jagoda2005,Casas06,casas2009}. Other parameters, like metallicity, were taken into account too, but the results on the mode identification did not change.\par

On the other hand we are aware that multicolour photometry identification predictions should be taken with care because rotation might have an influence on them \citep{pagoda02}. In particular, mode identification predictions depend on the angle of inclination of the star, and thereby on the rotation velocity (more exactly on the deformation of the star caused by rotation). Here, the uncertainties in the colour indices together with the large range of most probable inclination angles (cf. Fig~\ref{fig:inclination}) make angle-dependent identification predictions not feasible.\par

\begin{figure}
\begin{center}
 \includegraphics[width=0.45\textwidth]{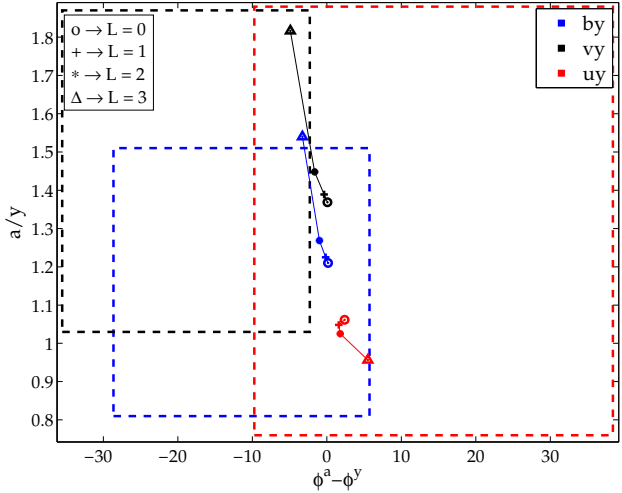}
 \caption{Phase shift vs. amplitude ratio diagram for the frequency of 21~\cd\ {{and a model with M~=~1.55~\msol, \alfa~=~0.5 and \ov~=~0.2}}. Every box correspond to a measure with its own uncertainties following the legend; band $y$ is the reference. Every symbol (o, +, $\ast$, $\Delta$) represents an \ele\ degree (0, 1, 2 and 3, respectively).}
 \label{fig:mixed mode}
\end{center}
\end{figure}

We illustrate the possible effect of rotation on the identification of modes using multi-colour photometry through the frequency that poses the most controversial mode-identification, namely the frequency near 21~\cd. As illustrated in Fig~\ref{fig:mixed mode}, the phase shift versus amplitude ratio diagram of this mode is quite peculiar. Band $b$ indicates that all the \ele's values are possible except \ele~=~3, whereas band v indicates the opposite.\par

In order to study the influence of rotation on the identification of that frequency, we studied the behaviour of one of the largest effects on the frequency due to rotation, the mode coupling or near degeneracy effects \citep[see ][]{Soufi98,suarez2006}. We computed a small grid of rotating models in the range of the physical parameters found in this work, but varying the rotational velocity over the most probable values obtained from the angle discriminant figure (Fig~\ref{fig:inclination}). Inclination values of $i\sim[35, 80]$ degrees result in rotational velocities between 200 and 125 \kms\ when considering models used in this analysis. These velocities imply ratios of $\Omega_{rot}/\Omega_{kep} = 0.47$ and 0.30 (with $\Omega_{kep} = \sqrt{G{M}/{R_{eq}^{2}}}$ and $R_{eq}$ being the equatorial radius), respectively. This range of values can be considered still below (but close to) the limit of validity of the perturbative approach, as stated by the test by \cite{Sua05altairII} and the comparisons analysis with the non-peturbative theory (see details in Sect.~\ref{subsec:rotation})\par

Asteroseismic rotating models were computed using the \cesam\ code, taking first-order effects of rotation into account. This is done by including the spherically-averaged contribution of the centrifugal acceleration, which is included by means of an effective gravity $g_{eff} = g-A_{c}(r)$, where $g$ is the local gravity, $r$ is the radius, and $A_{c} = 2/3 r \Omega^{2}$ is the centrifugal acceleration of matter elements. The non-spherical components of the centrifugal acceleration are included in the adiabatic oscillation computations \citep{Sua06pdrot}. Adiabatic oscillations were computed using the adiabatic oscillation code \filou\ \citep{SuaThesis,filou08}. This code provides theoretical adiabatic oscillations of a given equilibrium model corrected up to the second-order for the effects of rotation. These include near-degeneracy effects, which occur when two or more frequencies are close to each other. In addition, the perturbative description adopted takes radial variation of the angular velocity (radial differential rotation) into account in the oscillation equations.\par

We constrained the models to those fitting the frequency at $\sim21$~\cd\ within a frequency uncertainty up to the Rayleigh frequency. We used an heuristic method based on the best frequency fitting. In this method, we considered as {\emph{best models}} those with the lowest values of the $\chi^{2}$ distribution.\par

Except one, all the \emph{best models} found the mode as non-coupled with values of $(n,\ell,m) = (1,2,-1)$ or $(0,3,2)$. Only the model with also the lowest $\chi^{2}$ showed that the frequency could be a coupling between the modes $(1,1,0)$ and $(-1,3,0)$. In any case, the interesting point here is the value of the rotation for the best models rather than the identification itself. Due to the methodology, both the rotation values and the identification are product of the heuristic, but do not interfere each other.\par

The best models show rotational velocities around 200 and 130~\kms, which correspond to the extremes of the most probable range of inclination angles, i.e., 35 and 76 degrees, respectively. In order to find models with $i$ close the 62.5 deg (the minimum of the angle discriminant in Fig.~\ref{fig:inclination}) we had to relax the criteria of \teff, \logg\ from the spectroscopic uncertainty box to the photometric one (see Sec.~\ref{subsec:spec_parameters}) and frequency match error (for 21~\cd) to twice the Rayleigh frequency. In short, the variation of the inclination angle here does not affect the predictions of the rotating models for that frequency.\par

The results on the identification are shown in Table~\ref{tab:lpv}. The compatible values of the spherical degree for each frequency are listed in the column \ele. The ambiguity of the 21~\cd\ mode is not clarified by the rotating models study, which identifies that frequency as an $\ell = 2$ or 3 multiplet or a possible coupled mode, in contrast to the spectroscopic results, predicting an $\ell = 1$ mode. Note anyway, that the spectroscopic identification does not take high rotation into account, i.e., the star is considered spherical. This is solved in our perturbative approach for the oscillation frequencies which takes into account the deformation of the star due to rotation. In any case, the mode identification explained here must be taken carefully since the perturbative approach is close to the limit of validity for \starb\ (see Sect~\ref{subsec:rotation}).\par

%-fig-%
%-fig-%

%--------------------------------------------------------------------------------------------------%
\section{{{Best fitting model}}\label{sec:uncertainty}}
%--------------------------------------------------------------------------------------------------%

We used the information of the physical parameters and the mode identification derived in this work to search for the best representative models of \starb, i. e., those that fulfil all the constraints derived from the observations. A new and denser grid of non-rotating models where computed to this aim.\par

We computed the range [1.25, 2.20]~\msol\ in mass with a step of 0.01 and the range [-0.52, 0.08]~dex in \feh\ with a step of 0.2, covering the whole \ds\ instability strip. We used the same codes as in the previous section: \cesam\ and \graco\ for the main-sequence non-rotating equilibrium and oscillation models respectively, {{and again non-adiabatic frequencies were computed}}. The models included Eddington atmospheres and the same values for the convective parameters \alfa\ (0.5, 1.0 and 1.5) and \ov\ (0.1, 0.2 and 0.3). We only considered the range of \ele~=~[0, 3], because the visibility of the modes in the integrated light decreases with the spherical degree \citep{dziembowski1977}. The frequencies with the highest amplitudes in the \corot\ observations will be mainly low order modes so the range of computed \ele's must be enough to describe them.\par

Around 500,000 models were computed. This grid includes all the models lying in the \starb\ uncertainty box (1~$\sigma$) obtained from the spectroscopic observations. To efficiently handle this large amount of models we took advantage of VOTA \citep[Virtual Observatory Tool for Asteroseismology, ][]{suarez2013}, a tool designed to easily handle stellar and seismic models, analyse their properties, compare them with observational data and find models representative of the studied stars.\par

We selected then the models fulfilling the spectroscopic 1~$\sigma$ uncertainty box determined in this work. A total of 426 models fulfilled \teff, \logg\ and \feh, which led to the following physical parameters: M~=~[1.49, 1.58]~\msol, R~=~[1.50, 1.73]~\rsol, L~=~[6.47, 8.92]~\lsol, \rhom~=~[0.43, 0.62]~\gcm\ and age~=~[826, 1306]~Myr. The value of -0.12 in metallicity of our grid is the unique compatible with the spectroscopic observations.\par 

Finally, we applied the results to the mode identification. This included the discriminant obtained from the photometric observations about the convective efficiency: $\alpha\neq1.5$. The frequencies used for the discrimination in the models were: 18.131, 23.152, 23.192, 24.122 and 26.955 \cd\ (209.85, 267.96, 268.43, 279.19 and 311.98 \muhz, respectively). These frequencies were selected because their mode identification is the only that could be compared with the computed modes of our grid of non-rotating models, i.e., these frequencies have $\ell\leq3$ modes and $m=-1$ or 1.\footnote{Although none of the frequencies were identified as $m=0$ in the LPV analysis, which is the only compatible with non-rotating models, this value is valid for the selected frequencies within the $\pm1$ uncertainty.} The frequency at $\sim21$~\cd\ was not used because it possible coupled nature.\par

To fit models, all the possibilities in the spherical degree (limited by the error bars and up to $\ell=3$) were taken into account as valid solutions\footnote{For example, for the 23.192~\cd\ frequency, we searched the closest mode with the possible \ele~=~[0, 1, 2].}. The result is that 21 models fulfilled all the constraints, giving the same values for the ($n$, \ele) pairs (see last two columns in Table~\ref{tab:lpv}). The characteristics of the models are: \teff~=~[7510, 7592]~K, \logg~=~[4.17, 4.19]~dex, R~=~[1.64, 1.69]~\rsol, M~=~[1.53, 1.56]~\msol, L~=~[7.82, 8.37]~\lsol, \rhom~=~[0.455, 0.487]~\gcm, age~=~[1046, 1186]~Myrs, \feh~=~$-0.12$~dex, \hc~=~[0.4398, 0.4767], \alfa~=~[0.5, 1.0], \ov~=~[0.1, 0.2]. The values for the mass and the radius are lower than those found in the spectroscopic analysis, although still within the uncertainties. This could be due to the slightly lower metallicity of our models. In addition, the limitation of using only the frequencies as $m=0$ modes is a source of inaccuracy.\par 

%--------------------------------------------------------------------------------------------------%
\subsection{Influence of rotation\label{subsec:rotation}}
%--------------------------------------------------------------------------------------------------%

The methods to determine physical parameters, to identify the modes and to calculate models presented in the previous sections do not properly account for rotational effects.\par

In the determination of the physical parameters, the rotational effects are taken into account through the division of the star into annular regions, each characterized by a single intensity spectrum, and reducing the flux spectrum to a one-dimensional sum \citep{valenti1996}. Therefore, rotational velocity is well determined but only as a maximum projected value (differential rotation has also an effect in the determination of the other observables). However, the resulting quantities of the effective temperature and surface gravity are a mean value of the integration of the stellar disk. For rapidly-rotating stars, such a value for describing the whole body is not reliable. Due to the flattening produced by rotation, these quantities differ from the pole to the equator. In that way, the formal errors derived from fitting the spectra are too small.\par

The mode identification efforts are less accurate. For \dss, the problem usually lies in the low number of frequencies observed from the ground, the unknown of the physical processes involved in the selection mechanisms and in the usually rapid rotation of these objects. The latter is the most complicated task to resolve, because even the perturbative methods reach their limit of validity to describe pulsations for typical rotational velocities of \dss. In recent years, some works have shed light on this problem, thanks to the full integration of the oscillation equations in 2D models \citep[see, for example, ][]{lignieres2006, reese2006}. The computations showed how the modes reach a new distribution within the oblate star.\par

This could impact spectroscopic line variations used in mode identification. The {\sc{FAMIAS}} code uses the Fourier parameter fit method \citep{zima2006} and the moment method \citep{balona1986a,balona1986b,balona1987,briquet2003} to carry a mode identification out. Both methods assume oscillations in the limit of linearity (sinusoidal variations) and slow rotation (neglecting second order rotational effects), questioning the obtention of a reliable mode identification for moderate to rapidly rotating stars. Consequently, the mode identification obtained in this work may be incorrect for some modes: those most affected by rotation. To know which frequencies are poorly-detected cannot be achieved with the tools and methods developed up to date.\par

Rotation effects are also non-negligible in the multi-colour photometry, as discussed in Sec.~\ref{sec:stromgren}. In this case, the value of the inclination angle of the star affects $\phi^{T}$, ${\delta T_{eff}}/{T_{eff}}$ and ${\delta g_{e}}/{g_{e}}$. This implies a dependency of the amplitude ratios and phase differences on the azimuthal order \citep{pagoda02,Casas06}. Pertubative methods are able to compute these implications for moderately rotating stars, but the extreme cases require a special treatment. \citet{reese2012}, using their code to fully integrate the oscillating modes, have presented a method to compute amplitude ratios in the Geneva photometric system. Although this is a promising result, their computations did not include non-adiabatic effects, which are essential for a true determination of the frequencies' amplitudes.\par

In addition, the star is no longer spherical and some quantities, like the radius, might not be determined with the direct comparison to 1D non-rotating models (the equatorial radius is different from the polar radius). These facts introduce a strong uncertainty and inaccuracy in the physical characteristics derived for \starb\ in the previous sections.\par

Despite these problems, the non-perturbative computations have shown that some structures, like the large separation and the rotational splitting, are still identifiable in the frequency distribution \citep{lignieres2006, lignieres2010}. This is the reason for searching for patterns and structures within the periodogram of \starb.

%--------------------------------------------------------------------------------------------------%
\section{Quasiperiodic patterns within the frequency spectrum\label{sec:patterns}}
%--------------------------------------------------------------------------------------------------%

To search for periodicities in the frequency set of \starb, we used the same Fourier analysis as in GH09.\par

An ideal periodic pattern is represented as a series of equally spaced Dirac deltas, i. e., a Dirac comb. The Fourier transform (FT) of a Dirac comb is another Dirac comb with inverse periodicity of the original one. So, the more the FT resembles a Dirac comb, the more confident we are that the frequency spectrum has a periodicity.\par
A realistic case can be represented as a Dirac comb, if our function is periodic, multiplied by a rectangular function (because the number of frequencies is finite):

%--eqn--
\begin{eqnarray}
 \mathrm{F}(\nu) = \Delta_{T}(\nu) \cdot \sqcap (\nu).
 \label{eq:funcion}
\end{eqnarray}
%--eqn--

Where $\Delta_{T}(\nu)$ is the Dirac comb function with a periodicity of $T$ and $\sqcap (\nu)$ is the rectangular function. Then, the FT is:

%--eqn--
\begin{eqnarray}
 \mathcal{F}(t) = \int^{\infty}_{\infty} \Delta_{T}(\nu) \cdot \sqcap (\nu) \cdot \mathrm{e}^{-2\pi \mathrm{i}\nu t} \mathrm{d}\nu = \Delta_{1/T}\ast\mathrm{sinc}(t).
 \label{eq:FT}
\end{eqnarray}
%--eqn--

Due to the rectangular function, the form of the FT is not a perfect Dirac comb, but a convolution with a $sinc$ function. The side lobes originated by the $sinc$ function hinder the identification of the periodicity. In addition, other characteristics of a true frequency spectrum make the identification difficult. A pattern with an inexact periodicity (quasiperiodic), where frequencies do not belong to the pattern or some values are missing, would produce broadened peaks, spurious peaks (``noise" in the FT) and less powerful submultiples.\par

\begin{figure}
\begin{center}
 \includegraphics[width=0.49\textwidth]{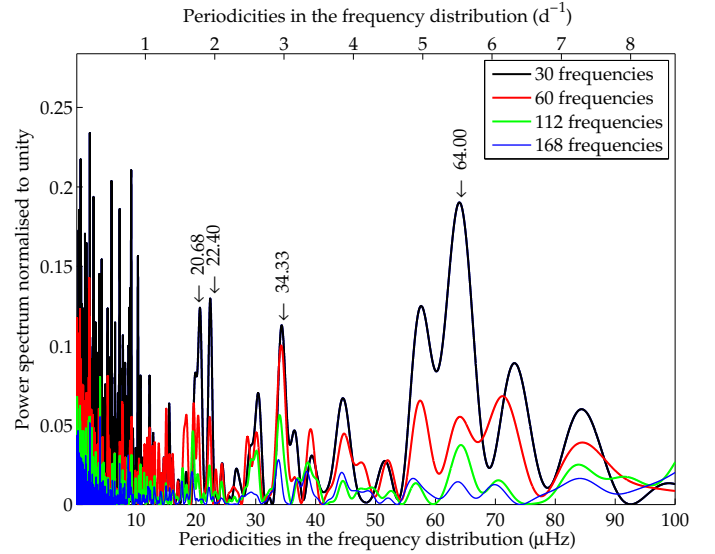}
 \caption{Fourier transform of various subsets selected by amplitude. The blue, green, red and cyan lines correspond to the FT calculated from the subset including the first 30, 60, 112 and 168 (all the peaks except the close ones) highest amplitude frequencies, respectively. The peak corresponding to the large separation (64~\muhz) and its sub-multiples are labelled (see text).}
 \label{fig:FT}
\end{center}
\end{figure}

We applied the method to the frequency spectrum of \starb. In our calculations, we neglected the amplitudes of the frequencies to avoid biases coming from the unknown mode selection. The resulting FT is given in Fig.~\ref{fig:FT}. We selected subsets of frequencies, namely subsets with the 30, 60, 112 highest frequencies, and a subset including all of them, for which we did not included the close peaks. Each subsequent subset contains the previous one. The main assumption we made is that the periodicity is due to frequencies with low \ele. The visibility of the modes decreases approximately as $\ell^{-2.5}$ or $\ell^{-3.5}$, depending whether the \ele\ degree is odd or even \citep{dziembowski1977}. This assumption is supported by the LPV analysis (see Sec.~\ref{subsec:lpv}), despite the possible inaccuracy of the mode identification method. The periodicity appears, above all, in the FTs of the subsets including a relatively low number of frequencies (30 and 60). We calculated the FT and presented the result in frequency scale, to obtain a value of the periodicity in \muhz.\par

%-fig-%
%-fig-%

We identify a peak as the periodicity of a pattern when the value and its sub-multiples\footnote{The periodicity of the Dirac comb resulting from the FT is k/T, where k is an integer. In the inverse scale of the plot, we will see T/k.} are found. As illustrated in Fig.~\ref{fig:FT}, we identified a pattern with a periodicity of 64~\muhz\ (5.53~\cd) in the frequency set. Any of the peaks we used to detect a pattern do not disappear when the number of frequencies involved in the FT is increased. And even when we do not use the {{12 peaks which might be combinations}}, the value of the periodicity does not change appreciably. These facts assure that the pattern is real.\par

%--------------------------------------------------------------------------------------------------%
\subsection{Large separation or rotational splitting structure?\label{subsec:large separation}}
%--------------------------------------------------------------------------------------------------%

The main question arising from the result in the previous section is the origin of the periodic pattern. Usually, the presence of a periodicity has been associated to the so-called large separation structure, that is common in frequencies that have reached the asymptotic regime, like those in the solar-type pulsators. However, it is not expected to appear in the regime where the \dss\ pulsations locate.\par

GH09 proposed that the origin of the periodicity could be linked to the large separation defined as usual: $\Delta\nu_{\ell}=\nu_{n+1,\ell}-\nu_{n,\ell}$, although out of the asymptotic regime. Fig.~4 of GH09 showed a plot of $\nu_{n,\ell}$ vs. \Dnu\ for a non-rotating model. \Dnu\ increases with frequency within the typical range of observed frequencies in \dss: $\nu\leq1000$~\muhz. However, within the interval [100, 600]~\muhz\ (this interval depends on the physical characteristics of the star and on the age), \Dnu\ comes to a halt and most of its values are confined within less than 4~\muhz. Although it is not a constant function, the range of variation is small enough to consider it as a roughly quasiperiodic regime and the FT would be able to detect it.\par

On the other hand, the origin of the periodic pattern has been also associated with the rotational splitting, which, for low rotation rates, forms multiplets with a periodicity equal to the rotational velocity. However, for large rotation values, multiplets become irregular \citep[see ][]{Soufi95,Sua06pdrot} breaking the periodic structure. Nonetheless, in a recent work, \citet{lignieres2010} searched for patterns within the frequency set computed with the non-perturbative method. To do that, they considered a visibility function for the computed frequency set and calculated its autocorrelation. They found, for some configurations of the inclination angle of the stellar rotational axis, a peak in the autocorrelation function corresponding to twice the rotational splitting frequency.\par

To disentangle between a periodic pattern coming from a large separation structure or from the rotational splitting, we have to analyse the case of \starb\ carefully. First, we have to note that the FT, as it is defined, is more sensitive to periodicities rather than other kind of spacings, such as a preferred randomly-distributed frequency difference. In the asymptotic regime, the large separation has a comb structure and should be easily discovered with the FT. The large separation structure is even conserved with high rotational velocities, because the structure is led by the so-called island modes \citep[this is the new distribution that appears when non-perturbative theory is applied to pulsation in fast rotators, ][]{lignieres2006}.\par

Concerning rotational splitting for the high velocity regime, modes containing the rotational information are the so-called chaotic modes, which are irregularly distributed in the frequency spectrum as the results of the studies based on the ray mechanics formalism showed \citep{lignieres2006}. One should not expect to find regular patterns in it. Despite that, they have specific statistical properties, which can be detected with an autocorrelation function. This is the case of the $2\Omega$ pattern showed in \citet{lignieres2010}, a randomly distributed frequency difference, which is, however, hardly identifiable using a FT.\par

In addition, using the physical parameters derived in the spectroscopic analysis for \starb, the rotational splitting should be 18.07~\muhz. Twice of this value is only a half of that found in the observed frequency set (64~\muhz). But even taking into account the uncertainties on the determination of the inclination angle (i. e., discriminant value $\pm$0.02) and the worst favourable case for the radius of the star that we derived from the models, the rotational splitting would be, at the most, $\sim$~29~\muhz.\par

%-fig-%
\begin{figure}
\begin{center}
 \includegraphics[width=0.49\textwidth]{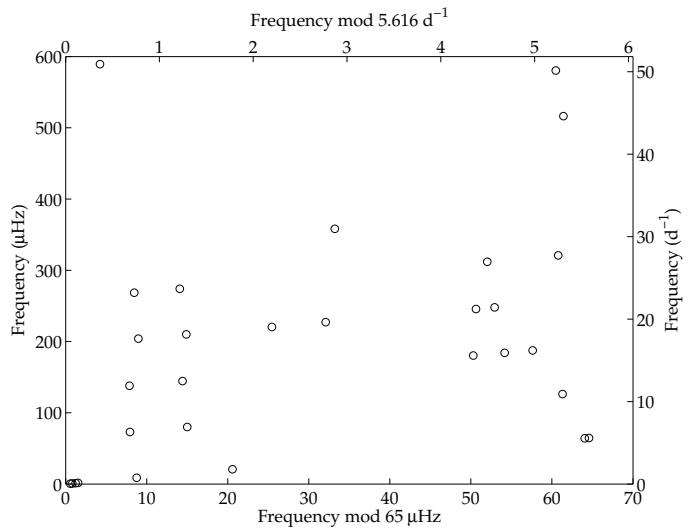}
 \caption{{{Echelle diagram for the 30 frequencies of highest amplitude of the \starb\ frequency spectrum using a periodicity of 65~\muhz.}}}
 \label{fig:echelle}
\end{center}
\end{figure}
%-fig-%

All this led us to think that the pattern found in \starb\ is a consequence of the large separation. But, in order to define definitively the structure of the pattern, we constructed an Echelle diagram, which can be seen in Fig.~\ref{fig:echelle}. In this plot, the 30 frequencies of highest amplitude of the \starb\ spectrum has been used, because they contain the information of the periodicity, and a spacing of 65~\muhz\ has been used. The frequencies involved in the pattern are clearly differentiated. Indeed, there are two groups of frequencies, placed at the left of the plot, which have an almost identical structure to the solar-like stars case. The main difference is the range in which the pattern appears, far from the asymptotic regime but around the fundamental radial mode value.\par

There is no doubt that the frequency periodic pattern (indeed, two patterns) for this object comes from a large separation-like structure. The question now is: is it possible to extract some information related to the interior of the star using this periodicity?

%--------------------------------------------------------------------------------------------------%
\section{The large separation as an observable\label{sec:constrain}}
%--------------------------------------------------------------------------------------------------%

There exists a previous attempt to use the large separation in \dss\ as an observable. \cite{breger2009} used a histogram of differences to search for periodicities. They found a periodic pattern in three \ds\ stars and systematized a method to reduce the uncertainties on the determination of the physical parameters. However, their method had two main problems. First, the frequency of the lowest unstable radial mode must be identified. This corresponds to a frequency within the cluster of modes at the lowest frequencies detected from the ground. It is a cluster because of the assumption of trapped modes. But space observations of \dss\ showed that in many cases it is difficult to identify a cluster with the lowest frequency, as can be seen in Fig.~\ref{fig:frequencies} and with \stara\ (GH09). Second, they took into account only the radial modes. As shown in {{Fig.~4 on GH09}}, the periodicity, in non-rotating models and in models with rapid rotation \citep{lignieres2006}, is not only formed by the modes with \ele~=~0. Even more, the large separation is better defined by the non-radial ones (until the avoided-crossing effects appear).\par

The aim of this work is to use the large separation found in the previous sections combined with that presents in the non-rotating models to obtain some information about the internal structure of the star. This method is supported by the fact that, although the value of the frequencies globally decreases when rotation increases and, so, the large separation, \Dnu\ is reasonably conserved up to $\Omega/\Omega_{K}\sim0.4$ (F.~Ligni\`{e}res \&\ D.~Reese, private communication). Because \starb\ rotates around or below this limit, we can suppose that the difference in the large separation between the rotating and non-rotating case is included in the error bars of \Dnu.\par

\begin{figure}
\begin{center}
 \includegraphics[width=0.49\textwidth]{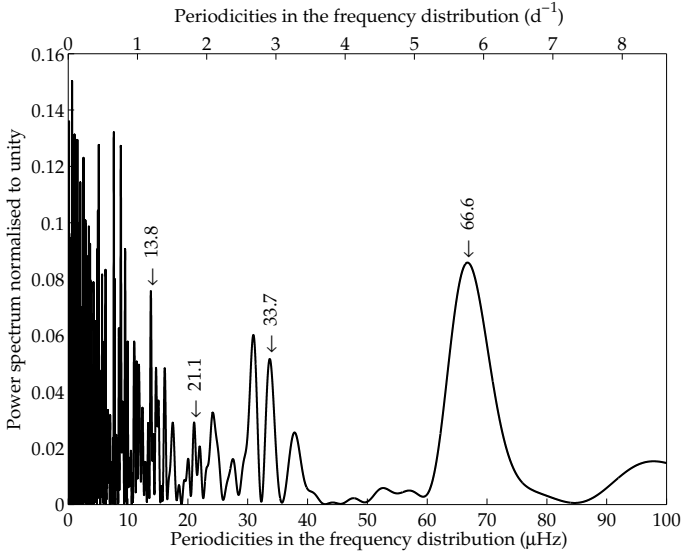}
 \caption{Fourier transform, as explained in Sec.~\ref{sec:patterns}, of a non-rotating model (M~=~1.55~\msol, \feh~=~-0.12, \alfa~=~0.5 and \ov~=~0.2) representative of \starb. A periodicity of $\sim67$~\muhz\ is clearly identifiable. Note that the FT was computed using only frequencies below 600~\muhz\ (51.84~\cd).}
 \label{fig:large separation}
\end{center}
\end{figure}

In Fig.~\ref{fig:large separation}, we applied the FT to the frequencies below 600~\muhz\ (to match the observations) of a non-rotating model representative of the star, i. e. fulfilling the spectroscopic physical parameters. A similar structure to that found in the observations can be seen, with the main peak plus the sub-multiples, resulting in a periodicity of $\sim67$~\muhz. Following this procedure, the models and the observations are directly comparable.\par

%-fig-%
%-fig-%

However, not all the physical quantities obtained from this comparison are reliable. As discussed in Sec.~\ref{subsec:rotation}, the physical parameters could differ from the slow-rotation case to the fast-rotation one. Fortunately, the work by \citet{reese2008} suggested that the large separation for the case of rapid rotators is still related in a simple way to the mean density of the star. And, even more, that the large separation to mean density ratio is roughly constant for all the rotation rates they computed. These results gave us the opportunity to determine the mean density of \starb\ using our grid of non-rotating models, even when the other quantities, such as the radius or the mass, could not be completely trusted.\par

As a starting point, we adopted the parameters \teff, \logg\ and \feh\ from {{the spectroscopic analysis: \teff~=~$7555\pm50$~K, \logg~=~$4.21\pm0.05$~dex and \feh~=~$-0.08\pm0.10$~dex (cf. Sec.~\ref{subsec:spec_parameters}).}} We used the non-rotating models of our grid covering the entire $1\sigma$ uncertainty box and looked for those with the same \Dnu\ (in the \ds\ pulsation regime) as that found in the observations. For simplicity, in the models, a mean value of \Dnu\ is calculated using only $p$~modes, because the large separation is a property of the acoustic modes, and within the observed range (until {{$n=8$}}), to be consistent with the observations. {{The observed \Dnu\ value was taken as an average of the frequency differences from the two columns that form the large separation pattern in the Echelle diagram (Fig.~\ref{fig:echelle}). Hence, $\Delta\nu=65\pm1$\muhz, with the uncertainty being the standard deviation.\par

For comparison, Table~\ref{tab:cajasfinal} summarizes the range in the physical parameters achieved by the classical methods (photometric and spectroscopic determination of \teff, \logg\ and \feh) and adding the use of the large separation. We also included the result obtained in the multi-colour analysis: \alfa~$\leq1.0$. As can be seen, with the large separation we obtained a precision in the mass, the radius, surface gravity and, above all, in the density of the star never reached before. As previously discussed, the results in the radius and mass should be taken with care. It is important to note that the large separation could not reduce the uncertainty in \teff. In addition, the uncertainties are a bit constrained by the fact that our grid did not scan enough of the space defined by the physical parameters (steps were too large).}}\par

%--table--
\begin{table}
   \caption{Table summarizing the range of validity of the physical quantities of \starb\ obtained using the different methods discussed in this work.}
\tiny
  \begin{center}
   \renewcommand{\arraystretch}{1.2}
   \begin{tabular}[ht!]{cccc}
   \hline
   \hline
   Method & Photometric box & Spectroscopic box & Using \Dnu\ \\
     \hline
     \teff\ (K) & $[7427, 7847]$ & $[7505, 7605]$ & {{[7505, 7605]}} \\
     \logg & [3.82, 4.25] & [4.16, 4.26] & {{[4.21, 4.23]}} \\
     \feh & [-0.32, 0.08] & {{[-0.18, +0.02]}} & {{[-0.18, +0.02]}} \\
     \alfa & [0.5, 1.5] & {{[0.5, 1.0]}} & {{[0.5, 1.0]}} \\
     \ov & [0.1, 0.3] & [0.1, 0.3] & {{[0.1, 0.3]}} \\
     M ($\mathrm{M}_{\odot}$) & $[1.35, 2.20]$ & $[1.49, 1.58]$ & {{[1.50, 1.53]}} \\
     R ($\mathrm{R}_{\odot}$) & [1.46, 3.02]  & [1.50, 1.73] & {{[1.55, 1.61]}} \\
     L ($\mathrm{L}_{\odot}$) & [5.87, 30.94] & [6.47, 8.92] & {{[6.8, 7.50]}} \\
     $\bar\rho$ (\gcm) & [0.113, 0.61] & [0.43, 0.62] & {{[0.51, 0.57]}} \\
     Age (My) & [434, 2244] & [826, 1306] & {{[926, 1206]}} \\
     $\mathrm{X_{c}}$ & [0, 0.7373] & [0.4105, 0.5676] & {{[0.4835, 0.5352]}} \\
     \hline
     \hline
   \end{tabular}
 \end{center}
 \label{tab:cajasfinal}
\end{table}
%--table--

The most important advantage of the method described above is that no assumption about the rotation of the star is needed to derive an accurate value of the mean density, a fundamental quantity, for example, in the characterization of a planetary system. The main assumption was about the visibility of the oscillation modes and the physical parameters taken as starting point in the analysis. In the work we present here, we have used the values of the high quality spectroscopic measurements, obtaining a value for \rhom\ with an uncertainty of 6\%. Nonetheless, even if the parameters from the photometry were used instead and no information about the \strom\ photometry was used, i. e., about the non-validity of \alfa=1.5, the resulting mean density would be of \rhom~=~[0.48, 0.59]~\gcm. This is a major improvement, because we obtain a better measurement of the mean density of the star than that achieved by any spectroscopic observation.\par

%--------------------------------------------------------------------------------------------------%
\subsection{Towards a reliable mode identification using the periodicity\label{subsec:mode ID}}
%--------------------------------------------------------------------------------------------------%

At this point we wondered if it is possible to extract more information from the large separation pattern. Mode identification is feasible in solar-like stars thanks to the Echelle diagram, using the structure of the pattern and its well known distribution with the spherical degree. A direct comparison between one of these diagrams for the model and for the observation, gives a direct identification of \ele. This is a consequence of the asymptotic regime, but it is not the case for \dss. Figure~\ref{fig:echelle} shows that the frequencies comprising the two columns pattern are below 300~\muhz. Plotting an Echelle diagram for our non-rotating models did not give this structure, only a straight distribution for some modes, which form the large separation comb used in the previous section.\par

We moved to our grid of rotating models searching for a representative one of \starb. Following the same procedure of selecting models fulfilling the spectroscopic parameters, we selected then those with a large separation within the uncertainties derived in the previous section ($65\pm1$~\muhz). The large separation, in this case, has been calculated as an average value of the frequencies with \ele~=~1 up to $n=8$, in order to match the observations (frequency range below $\sim600$~\muhz). \ele~=~1 modes are the best determined in the second-order perturbative approximation compared to the values computed with non-perturbative methods \citep{lignieres2006}. This indicates that some reliable conclusions can be obtained from the distribution of the spherical degrees observing our rotating models.\par

The 6 final models selected cover the uncertainty range of the inclination angle, meaning rotational velocities from $\sim125$~\kms\ up to $\sim200$~\kms. Coherently with the predictions by \cite{reese2008}, the mean density of these models discriminated by large separation is [0.54, 0.58]~\gcm, which is in the same range as for the non-rotating ones (see Table~\ref{tab:cajasfinal}). The masses, however, vary from 1.55~\msol\ to 1.61~\msol, slightly higher than that obtained in the previous section.\par

\begin{figure}
\begin{center}
 \includegraphics[width=0.49\textwidth]{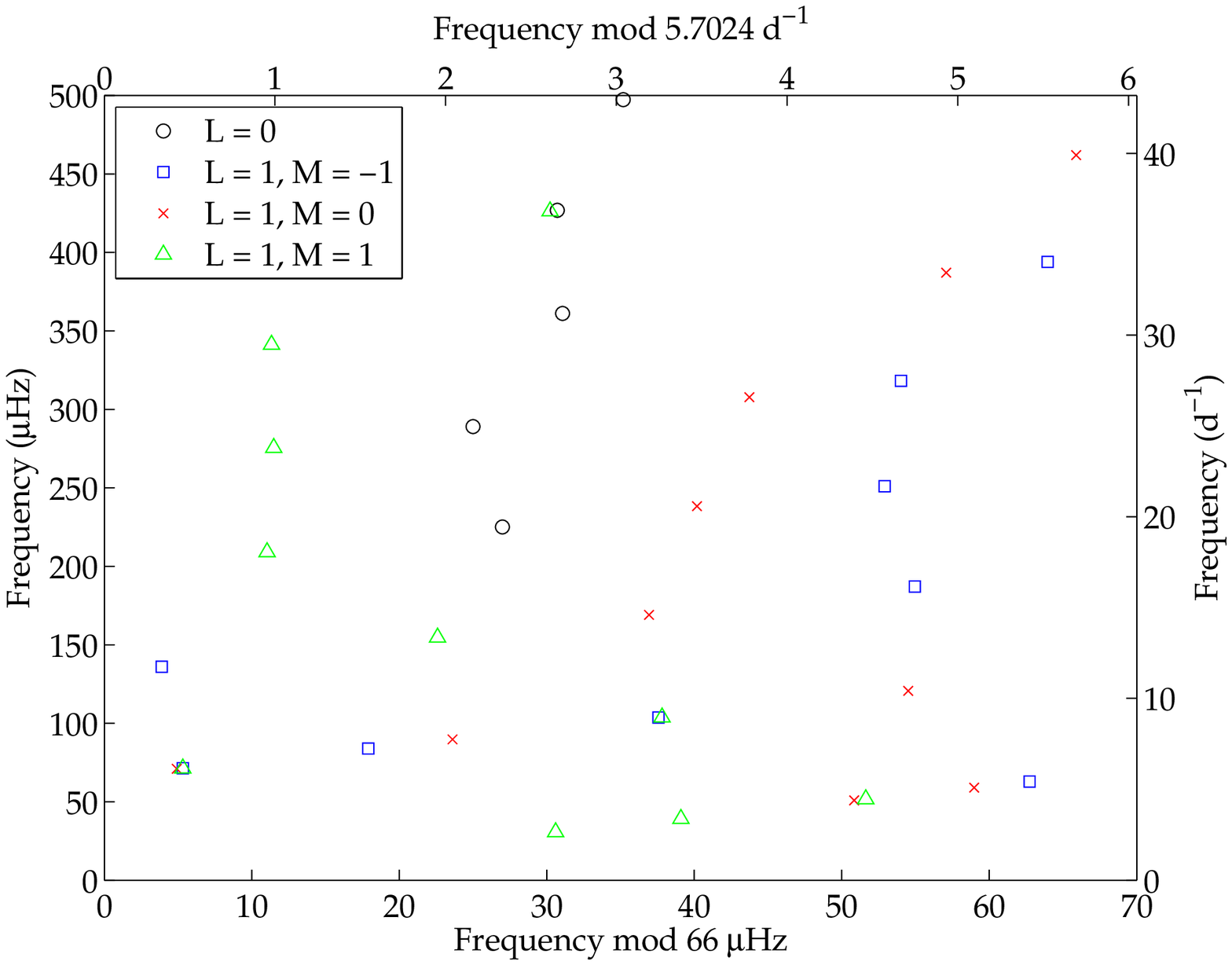}
\caption{Echelle diagram for a 1.61~\msol\ rotating model of \starb, with $\sim135$~\kms. Each couple ($\ell=0$; $\ell=1, m=-1$; $\ell=1, m=0$; $\ell=1, m=1$) is depicted with a different colour (black, blue, red and green, respectively) and a symbol ($\bigcirc$, $\square$, X and $\vartriangle$, respectively). A structure similar to the large separation found in the observations can be seen for some of these pairs between 200-350~\muhz. This structure is mainly formed by the \ele~=~1 modes.}
 \label{fig:echelle teorico}
\end{center}
\end{figure}

We applied the FT procedure to the models in order to find a periodicity and used it to construct an Echelle diagram. An example can be seen in the Fig.~\ref{fig:echelle teorico}. We only used the lowest spherical degrees (\ele~=~0, 1), assuming the same visibility of the mode as in Sec.~\ref{sec:patterns}, and frequencies below 500~\muhz\ to select the range of the observed Echelle. All the plots showed the same structure: the \ele~=~0 modes aligned with the \ele~=~1, but only with $m=-1$ and $m=1$, and the \ele~=~1, $m=0$ modes form another structure (indeed, also becoming a pattern, but with a slightly higher periodicity, typically $\sim5$~\muhz\ more). The difference between the azimutal orders $m=-1$ and $m=1$ appeared independent from the model, although the $m=0$ values showed a displacement with rotation. The most different behaviour was found in the model with the highest rotational velocity.\par

%-fig-%
%-fig-%

It is not feasible to say that we are able to do a mode identification using these models or even to say that the observed patterns are \ele~=~1 modes. Although the pattern is seen in the models, it is clear that some gaps exists between the observations and the theory. First, the columns in the Echelle diagram of the \corot\ frequencies are closer than any pair of combs in the models. Second, the frequencies in the observations are more tightly distributed. And, finally, the lowest frequencies in the pattern of the observations lie below 100~\muhz, which is lower than the values of the patterns in the models.\par

Maybe these differences could be explained because frequencies at high rotation rates tend to decrease with respect to their non-rotating values \citep{lignieres2006}. Nonetheless, this range below 100~\muhz\ typically corresponds to $g$ modes. Any recent works related to pulsations in highly rotating stars, never showed the combined analysis of the pressure and gravity modes. It seems, looking at the observations, that a new regime is also reached at these velocities around the fundamental radial mode. It could be useful to note that at 30\%\ of the Keplerian limit, an interesting reorganization of $p$ modes occurs \citep[Fig. 4 in ][]{reese2006}. It is not an exaggeration to think that this might be the case but with the additional effect in the $g$ modes. This study has not yet been made; therefore, a mode identification is not possible at this point.\par

Regardless of this last conclusion, the searching for structures within the frequency spectra of \dss\ seems to be one of the best ways to finally achieve a fully understanding of the interior of this complex pulsators. The recent development of the non-perturbative theory is also needed but, because computations are, up to now, highly-time consuming, the comparison with non-rotating models, as showed in this work, could be the way to characterize a large sample of A-F type stars as those provided by the \kepler\ satellite.\par

%--------------------------------------------------------------------------------------------------%
\section{Conclusions\label{sec:conclusions}}
%--------------------------------------------------------------------------------------------------%

In the present work, we used the most advanced methodology available to study the \ds\ star \starb. In particular, we analysed the \corot\ light curve and performed new high resolution spectroscopic and multi-colour photometric ground-based observations. We carried out an in-depth study of this pulsator, comparing the result from every single method and using all the information to try to derive the most accurate physical parameters of its interior.\par

From the \corot\ light curve, we extracted 185 frequencies above the limit of sig~=~10 within the range from a value close to zero up to about 77~\cd, i. e., 900~\muhz. This limit translates to 8 ppm for the smallest peak. We found that 17 of these peaks were located closer than the limit resolution to other. In addition, 37 more peaks were identified as possible combinations, although 25 of them seemed to be only coincidences. In summary, 160 peaks could be considered as independent frequencies. This number is significantly smaller than previous \dss\ observed by \corot, but we have to take into account the shorter time baseline.\par

From the spectroscopic observations, we derived the physical parameters, including the inclination angle and the rotational velocity. We extracted 18 frequencies and performed a mode identification. Most of them corresponded to the modes extracted from the \corot\ light curve and one appeared to be a double peak not resolved in the \corot\ frequency set.\par

In addition, the three highest peaks of the 160 CoRoT frequencies were also detected from ground-based $uvby$ photometry. The mode identification from multi-colour photometry is in good agreement with that from high-resolution spectroscopy. For the lowest amplitude peak we found a discrepancy, whose detailed analysis using rotating models indicated an $\ell=2$ or 3 non-coupled mode or a coupling between modes $\ell=1$ and 3, in contrast with the spectroscopic value, $\ell=1$. Another important result concerns to the \alfa\ parameter of the convention theory: $\alpha<1.5$, which is lower than that usually used for intermediate mass stars.\par

We tried to constrain the modelling of the star using the information from the proposed mode identification. 21 models were found to fulfil all the constraints and its physical parameters were slightly lower from those found in the spectroscopic analysis, although within the uncertainties. Nevertheless, these results might not be very accurate, because, for instance, models did not account properly for fast rotation. In particular, we need non-perturbative models that include properly the rotational effects, although they are very time-consuming. On the other hand, the fast rotation amplifies the differences between stellar and observer frames. Two modes with different $m$ could be unresolved in the observer reference frame. In such a case it could be very difficult to correctly decipher the resulting amplitude and phase diagrams and provide a reliable mode identification. In addition, the LPV method neglects the second order rotational effects, so it is not prepared to describe the new order of the modes showed by non-perturbative computations. These facts make difficult  the comparison between theory and observations and, consequently, the extraction of accurate information about the stellar interior.\par

As in previous cases (\stara, GH09, and HD\,50870, \citealt{mantegazza2012}), we searched for periodic patterns within the frequency set of \starb\ expecting that a relative measurement could be used to infer useful information from the spectrum. We found a periodicity of $\sim$64~\muhz\ (5.53~\cd). Thanks to the estimate of the inclination angle obtained from the analysis of the line-profile variations and an Echelle diagram clearly showing a structure, we were able to discard the rotational origin of the pattern and identify it as a large separation. We studied the possibility of using \Dnu\ as an observable, discriminating non-rotating models of a grid with the objective of reducing the uncertainty in the mean density of the star. Through the Echelle diagram, a tight value of the large separation was obtained: $\Delta\nu=65\pm1$\muhz. Using this value, we found that \rhom~=~[0.51, 0.57]~\gcm, which supposes an uncertainty of 6\%, never reached before for a non-binary \ds\ star. The main advantage of this method is that no initial assumptions related to the rotation are needed to find the periodicity from the observations because its value depends only slightly on the velocity of the star. \par

Moreover, comparing the observed Echelle diagram to other rotating models computed using second order terms in the perturbative theory, we found a similar structure of the large separation, formed mainly by the \ele~=~1 modes with $m$~=~-1, 1. The pattern observed also seems to include some gravity modes, which was unexpected and never described in the literature. Nonetheless, although the models could not be used in this case to perform a mode identification, we showed that it might be feasible using the new modelling techniques, and that the large separation pattern is a useful tool in this task even for \dss. Further investigations on the Echelle diagrams of theoretical models should be done to elucidate the organization observed in the pulsating modes.\par

All these results demonstrate the usefulness of the large separation in \dss. It will allow the determination of the mean density of one object with this kind of pulsations regardless of its rotational velocity (this assertion will have to be confirmed for every rotation rate). This is a major improvement for the study and characterization of, for example, planetary systems hosting A-F type stars and an affordable method for the precise study of \ds\ pulsators.

%--------------------------------------------------------------------------------------------------%
\acknowledgements {The authors thank the anonymous referee for useful comments which allowed them to improve the paper. The authors wish to thank L.~Mantegazza for having performed important parts of the spectroscopic analysis. The authors wish to thank John Telting for useful comments on an advanced draft of the paper. AGH wishes to thank S.~Murphy for his carefully revision of the text. AGH was supported by grant BES2005-8478, under the project ``Participaci\'on espa\~nola en la misi\'on \corot'' (ESP2004-3855-C03-01), and is supported by grant SFRH/BPD/80619/2011, from FCT (Portugal). AGH acknowledges financial support from project PTDC/CTE-AST/098754/2008 from FCT (Portugal). AM acknowledges the funding of AstroMadrid (CAM S2009/ESP-1496) and the Spanish grants ESP2007-65475-C02-02, AYA 2010-21161-C02-02. JCS acknowledges support by the Spanish National Research Plan (grants ESP2010-20982-C02-01, AYA2010-12030-E). MR and EP acknowledge financial support from the Italian PRIN-INAF 2010 {\it Asterosesismology: looking inside the stars with space- and ground-based observations}. PJA acknowledges financial support from grant AYA2010-14840 of the previous Spanish Ministry of Science and Innovation (MICINN), currently Ministry of Economy and Competitiveness. KU acknowledges financial support by the Spanish National Plan of R\&D for 2010, project AYA2010-17803. ES and CR acknowledge support from the Spanish Virtual Observatory financed by the Spanish MICINN and MINECO through grants AyA2008-02156 and AyA2011-24052. This work was supported by GRID-CSIC Project (200450E494). This research benefited from the computing resources provided by the European Grid Infrastructure (EGI). This investigation was partially supported by the Junta de Andaluc\'\i a and the Spanish Direcci\'on General de Investigaci\'on (DGI) under project AYA2009-10394.}
%--------------------------------------------------------------------------------------------------%

%--------------------------------------------------------------------------------------------------%
\bibliographystyle{aa}

%--------------------------------------------------------------------------------------------------%

%-tab-%
\onecolumn
\begin{landscape}
\begin{center}

\tablefirsthead{%
   \hline
   \hline
   \multicolumn{1}{c}{Id} & \multicolumn{1}{c}{Freq. (\cd)} & \multicolumn{1}{c}{Freq. (\muhz)} & \multicolumn{1}{c}{Amp. ($mmag$)} & \multicolumn{1}{r}{Phase ($rad$)} & \multicolumn{1}{c}{Sig} & \multicolumn{1}{c}{S/N} & \multicolumn{1}{c}{rms} & \multicolumn{1}{c}{$\sigma_{f}$ (\cd)} & \multicolumn{1}{c}{$\sigma_{A}$ $(mmag)$} & \multicolumn{1}{c}{$\sigma_{ph}$ $(rad)$} & \multicolumn{1}{c}{Comments} \\
   \hline
}
\tablehead{%
   \hline
   \hline
   \multicolumn{1}{c}{Id} & \multicolumn{1}{c}{Freq. (\cd)} & \multicolumn{1}{c}{Freq. (\muhz)} & \multicolumn{1}{c}{Amp. ($mmag$)} & \multicolumn{1}{r}{Phase ($rad$)} & \multicolumn{1}{c}{Sig} & \multicolumn{1}{c}{S/N} & \multicolumn{1}{c}{rms} & \multicolumn{1}{c}{$\sigma_{f}$ (\cd)} & \multicolumn{1}{c}{$\sigma_{A}$ $(mmag)$} & \multicolumn{1}{c}{$\sigma_{ph}$ $(rad)$} & \multicolumn{1}{c}{Comments} \\
   \hline
}
\tablelasttail{%
\multicolumn{12}{l}{{{$^{7}$Full table is only available in electronic form at the CDS via anonymous ftp to cdsarc.u-strasbg.fr (130.79.128.5) or via http://cdsweb.u-strasbg.fr/cgi-bin/qcat?J/A+A/}}} \\}
\tablecaption{{{Frequencies from the \corot\ light curve for \starb. The frequency in \cd\ and \muhz, the amplitude, phase, spectral significance and S/N are listed, as well as the r.m.s. and the errors in the determination of the frequencies, amplitudes and phases. Last column contains the close frequencies (see text) and the possible combinations.}}}

\begin{supertabular}{l*{2}{r}c*{5}{r}{c}{r}p{3cm}}
\label{tab:frequencies} 

F1 & 23.19481516 & 268.45850880 & 6.2902 & -0.849284 & 7928.848 & 3384.993 & 6.000 & 0.00000376 & 0.0012 & 0.000186 &  \\
F2 & 26.95851157 & 312.01980984 & 5.1034 & -2.388499 & 11620.559 & 2462.099 & 4.029 & 0.00000464 & 0.0012 & 0.000229 &  \\
F3 & 21.42080448 & 247.92597778 & 2.0929 & -1.891001 & 9997.027 & 1083.607 & 1.779 & 0.00001131 & 0.0012 & 0.000558 &  \\
F4 & 27.71548633 & 320.78109178 & 0.9973 & 3.001985 & 7364.347 & 482.136 & 0.987 & 0.00002374 & 0.0012 & 0.001171 &  \\
F5 & 17.62250520 & 203.96418056 & 0.6038 & 0.109671 & 5757.060 & 261.153 & 0.691 & 0.00003922 & 0.0012 & 0.001935 &  \\
F6 & 50.15324453 & 580.47736725 & 0.3111 & 1.475902 & 2440.658 & 170.487 & 0.535 & 0.00007611 & 0.0012 & 0.003755 & F1+F2 = 50.1533 \\
F7 & 12.47882154 & 144.43080486 & 0.2462 & 1.336730 & 1851.490 & 111.131 & 0.488 & 0.00009619 & 0.0012 & 0.004745 &  \\
F8 & 18.13525291 & 209.89876053 & 0.2308 & 1.666664 & 1828.671 & 100.461 & 0.456 & 0.00010262 & 0.0012 & 0.005063 &  \\
F9 & 0.07030513 & 0.81371678 & 0.1720 & 2.908728 & 373.807 & 102.972 & 0.300 & 0.00013772 & 0.0012 & 0.006794 &  \\
F10 & 0.10569453 & 1.22331632 & 0.1694 & -3.129712 & 642.924 & 101.429 & 0.350 & 0.00013981 & 0.0012 & 0.006898 &  \\
\hline
F11 & 0.04950000 & 0.57291667 & 0.1506 & 2.639740 & 211.093 & 90.158 & 0.279 & 0.00015729 & 0.0012 & 0.007760 & Close to F9 \\
F12 & 50.91071289 & 589.24436215 & 0.1497 & 2.265424 & 878.604 & 84.535 & 0.426 & 0.00015818 & 0.0012 & 0.007804 & F1+F4 = 50.9103 \\
F13 & 5.53344382 & 64.04448866 & 0.1406 & -2.669389 & 862.550 & 75.119 & 0.413 & 0.00016839 & 0.0012 & 0.008308 & $|$F3-F2$|$ = 5.5377 \\
F14 & 19.62083260 & 227.09296991 & 0.1361 & 0.861799 & 815.144 & 62.480 & 0.400 & 0.00017400 & 0.0012 & 0.008584 &  \\
F15 & 44.61829076 & 516.41540231 & 0.1301 & -2.113708 & 829.737 & 69.705 & 0.389 & 0.00018208 & 0.0012 & 0.008983 & F1+F3 = 44.6156 \\
F16 & 15.57686071 & 180.28773970 & 0.1238 & -1.512670 & 796.042 & 54.442 & 0.377 & 0.00019134 & 0.0012 & 0.009440 &  \\
F17 & 30.95009515 & 358.21869387 & 0.1157 & -2.092364 & 693.349 & 61.490 & 0.367 & 0.00020468 & 0.0012 & 0.010098 & 2FS+3 = 30.9440 \\
F18 & 15.91208339 & 184.16763183 & 0.1057 & 1.134954 & 629.656 & 45.705 & 0.358 & 0.00022408 & 0.0012 & 0.011055 &  \\
F19 & 16.21017689 & 187.61778808 & 0.1002 & -2.305530 & 588.781 & 42.702 & 0.342 & 0.00023630 & 0.0012 & 0.011658 & $|$-[F4-(3FS+2)]$|$= 16.2005 \\
F20 & 19.04660919 & 220.44686563 & 0.0978 & -2.266460 & 590.737 & 43.663 & 0.335 & 0.00024219 & 0.0012 & 0.011948 &  \\
\hline
F21 & 11.91308048 & 137.88287593 & 0.0975 & 2.411291 & 614.601 & 44.605 & 0.328 & 0.00024284 & 0.0012 & 0.011981 &  \\
F22 & 0.04950000 & 0.57291667 & 0.0878 & 1.224443 & 502.250 & 52.591 & 0.316 & 0.00026965 & 0.0012 & 0.013303 & = F11 \\
F23 & 21.22309483 & 245.63767164 & 0.0821 & -1.580787 & 477.953 & 42.130 & 0.310 & 0.00028830 & 0.0012 & 0.014224 &  \\
F24 & 1.77854494 & 20.58501088 & 0.0783 & 2.976526 & 485.494 & 47.024 & 0.321 & 0.00030241 & 0.0012 & 0.014919 & $|$F3-F1$|$ = 1.7740 \\
F25 & 0.13230194 & 1.53127245 & 0.0782 & 0.210334 & 271.230 & 46.810 & 0.276 & 0.00030294 & 0.0012 & 0.014946 & Close to F10 \\
F26 & 6.30246373 & 72.94518206 & 0.0748 & -2.750323 & 442.737 & 38.480 & 0.305 & 0.00031669 & 0.0012 & 0.015624 & $|$F4-F3$|$ = 6.2947 \\
F27 & 6.91300419 & 80.01162257 & 0.0687 & 0.306557 & 363.045 & 34.224 & 0.296 & 0.00034481 & 0.0012 & 0.017011 &  \\
F28 & 23.19771104 & 268.49202593 & 0.0631 & -0.145173 & 269.311 & 33.938 & 0.286 & 0.00037548 & 0.0012 & 0.018524 & Close to F1 \\
F29 & 0.75789031 & 8.77187859 & 0.0591 & 2.852021 & 360.599 & 35.464 & 0.293 & 0.00040059 & 0.0012 & 0.019763 & $|$F4-F2$|$ = 0.7570 \\
F30 & 23.68167050 & 274.09340856 & 0.0562 & 1.538126 & 270.901 & 30.044 & 0.289 & 0.00042163 & 0.0012 & 0.020801 &  \\
\hline
F31 & 10.91275821 & 126.30507187 & 0.0475 & 0.591506 & 216.755 & 21.956 & 0.283 & 0.00049852 & 0.0012 & 0.024595 & \\
F32 & 5.57944634 & 64.57692523 & 0.0469 & -0.151677 & 194.272 & 25.016 & 0.273 & 0.00050451 & 0.0012 & 0.024890 & $|$F5-F1$|$ = 5.5723 \\
F33 & 0.21222393 & 2.45629549 & 0.0463 & -2.831574 & 213.580 & 27.710 & 0.281 & 0.00051176 & 0.0012 & 0.025248 &  \\
F34 & 20.62812578 & 238.75145579 & 0.0442 & 2.505928 & 193.343 & 22.072 & 0.271 & 0.00053526 & 0.0012 & 0.026407 &  \\
F35 & 17.40844864 & 201.48667407 & 0.0432 & -0.981306 & 163.984 & 18.612 & 0.265 & 0.00054801 & 0.0012 & 0.027036 &  \\
F36 & 25.09353708 & 290.43445694 & 0.0429 & 1.706041 & 193.911 & 21.976 & 0.269 & 0.00055187 & 0.0012 & 0.027227 &  \\
F37 & 0.16338575 & 1.89103877 & 0.0422 & -2.425021 & 177.171 & 25.286 & 0.267 & 0.00056082 & 0.0012 & 0.027668 & Close to F25 \\
F38 & 24.30838825 & 281.34708623 & 0.0398 & -0.936718 & 145.147 & 21.078 & 0.262 & 0.00059564 & 0.0012 & 0.029386 &  \\
F39 & 20.82386719 & 241.01698137 & 0.0379 & 0.066842 & 138.480 & 19.057 & 0.261 & 0.00062557 & 0.0012 & 0.030863 &  \\
F40 & 30.22708013 & 349.85046447 & 0.0376 & -1.169072 & 145.903 & 19.391 & 0.264 & 0.00062962 & 0.0012 & 0.031063 &  \\
\hline
\hline
\\

\end{supertabular}
\end{center}
\end{landscape}
%-tab-%

%++++++++++++++++++++++++++++++++++++++++++++++++++++++++++++++++++++++++++++++%
\end{document}